\documentclass[apj]{emulateapj}
\usepackage[colorlinks,linkcolor={blue},citecolor={blue},urlcolor={red}]{hyperref}
\usepackage{mathtools}
\usepackage{epsfig,graphicx,natbib,amsmath,amsfonts,amssymb}%,xfrac}
\usepackage{color}
\usepackage{multirow}
\usepackage{booktabs}
\usepackage{amssymb}
\usepackage{threeparttablex} % for "ThreePartTable" environment
\usepackage{booktabs} 
\usepackage{graphicx}
\usepackage{epstopdf}
\usepackage{longtable}
\usepackage{array} % for extrarowheight
\usepackage[dvipsnames]{xcolor}  

\newcommand{\RSFRL}{${\rm SFR}_{\rm 5Myr}/{\rm SFR}_{\rm 800Myr}$}

\newcommand{\re}{\Reff}
\newcommand{\msolar}{${\rm M}_\odot$}

\newcommand{\dindex}{D$_n$(4000)}

\newcommand{\hd}{H$\delta$}
\newcommand{\hda}{\hd$_A$}
\newcommand{\ewhda}{EW(\hda)}
\newcommand{\ha}{H$\alpha$}
\newcommand{\hae}{\ha}
\newcommand{\ewhae}{EW(\hae)}

\newcommand{\Reff}{{$R_{\rm e}$}}

\newcommand{\myemail}{\email{enci.wang@phys.ethz.ch}}

\defcitealias{Wang-19a}{W19}
\defcitealias{Wang-19b}{Paper I}

\shorttitle{The time variation of SFHs}
\shortauthors{Wang \& Lilly}

\graphicspath{{fig3/}}
\begin{document}

\title{The variability of star formation rate in galaxies: II. power spectrum distribution on the Main Sequence}

\author{
Enci Wang\altaffilmark{1},
Simon J. Lilly\altaffilmark{1}
} \myemail

\altaffiltext{1}{Department of Physics, ETH Zurich, Wolfgang-Pauli-Strasse 27, CH-8093 Zurich, Switzerland}

\begin{abstract}

We 
constrain the temporal power spectrum 
of the sSFR(t) of star-forming galaxies, using 
a well-defined sample of Main Sequence galaxies from MaNGA and our earlier measurements of the ratio of the SFR averaged within the last
5 Myr to that averaged over the last 800 Myr. 
We explore the assumptions of stationarity and ergodicity that are implicit in this approach. 
We assume a single power-law form of the PSD but introduce an additional free parameter, 
the ``intrinsic scatter'', to try to account for any non-ergodicity introduced from various sources. 
We analyze both an ``integrated'' sample consisting of
global measurements of all of the galaxies, and also 25 sub-samples obtained by considering five radial regions and five bins of integrated stellar mass.
Assuming that any intrinsic scatter is not the dominant contribution 
to the Main Sequence dispersion of galaxies,
we find that 
the PSDs have slopes between 1.0 and 2.0,  
indicating that the power (per log interval of frequency) is mostly contributed by 
longer timescale variations. 
We find a correlation between the 
returned PSDs and the inferred gas depletion times ($\tau_{\rm dep,eff}$) obtained from application of the extended Schmidt Law, in that
regions with shorter gas depletion times
show larger integrated power and flatter PSD.  Intriguingly, it is found that shifting the PSDs by the inferred $\tau_{\rm dep,eff}$ causes all of the 25 PSDs to closely overlap, at least in that region where the PSD is best constrained and least affected by uncertainties about any intrinsic scatter. 
A possible explanation of these results is the dynamical response of the gas regulator 
system of \cite{Lilly-13} to a uniform time-varying inflow, as previously proposed in \cite{Wang-19a}. 

\end{abstract}

\keywords{galaxies: general -- methods: observational}

\section{Introduction}
\label{sec:introduction}

Galaxies are usually separated into
two types according to their locations on the 
color-magnitude diagram or star formation rate-stellar
mass (SFR-$M_*$) diagram: star-forming (SF) galaxies and passive ``quenched" ones
\citep[e.g.][]{Strateva-01, Baldry-04, Bell-04, Blanton-05, Li-06, Faber-07,
Wetzel-12}. 
Typical SF galaxies have continuous 
on-going star formation and disk-like morphology. 
Although the great progress has been made in understanding the properties of SF galaxies in   
the last two decades, the temporal variability of the star formation rates of individual galaxies is still poorly understood.  

SF galaxies are located on a tight sequence on the SFR-$M_*$ diagram, both in the local Universe and 
up to redshift of 3 \citep[e.g.][]{Brinchmann-04, Daddi-07, Elbaz-07, Noeske-07, Elbaz-11}.  This tight sequence is known as the star formation Main Sequence
(SFMS).  The scatter of the specific SFR (sSFR) at a given mass on the SFMS measured from the observations varies between 0.2-0.4 dex
for different authors \citep[e.g.][]{Whitaker-12, Speagle-14, Schreiber-15, 
Boogaard-18}, depending in detail on the definition of the sample of SFMS galaxies, as well as on how the stellar mass and SFR 
are measured. The scatter of the SFMS depends on both the stellar mass and structural properties of galaxies. 
\cite{Wang-18b} found that the scatter of the SFMS depends strongly on the structure of 
galaxies, with more compact galaxies showing a larger range of sSFR. 
By using different SFR measurements, including H$\alpha$, UV and infrared emission, 
\cite{Davies-19} found that the scatter of the SFMS increases with increasing stellar mass at stellar masses greater
than $10^{9.2}$\msolar. 
Overall, there is no evidence for any strong redshift evolution of the scatter of the SFMS \citep{Whitaker-12, Speagle-14, Schreiber-15}. 
The evident stability of the SFMS (i.e. a scatter nearly independent of redshift) has been interpreted as the quasi-steady-state 
interplay between cold gas inflow, star formation and outflow \citep[e.g.][]{
Schaye-10, Bouche-10, Dave-11, Lilly-13, Tacchella-16, Wang-19a}.

Indeed, models based on this quasi-steady-state interplay (also known as gas-regulator model) has achieved great 
success in explaining the observational facts of SF galaxies.  For instance, \cite{Lilly-13} found that
their simple gas-regulator model should produce the observed fundamental metallicity relation, i.e. not only that the SFR of galaxies should be a second parameter in the mass-metallicity relation, but also that the $Z(M_*,{\rm SFR})$ relation should be epoch-independent \citep[e.g.][]{Mannucci-10, Lilly-13, Belfiore-19}. 

A key question is how galaxies vary over time on and around the SFMS, i.e. whether they vary their sSFR relative to the mean SFMS, and, if so, on what timescales. This question will be a focus of the current paper. 

Recently, \citet[][hereafter \citetalias{Wang-19a}]{Wang-19a} found that galaxies above (or below) 
the SFMS show higher (or lower) SFR surface density ($\Sigma_{\rm SFR}$) at all galactic radii with 
respect to the median $\Sigma_{\rm SFR}$ profile of typical SF galaxies. In addition, \citetalias{Wang-19a} also found that 
the dispersion of $\Sigma_{\rm SFR}$ (at a given relative radius in galaxies of similar stellar mass) decreases with increasing 
(inferred) effective gas depletion time - the effective gas depletion timescale being inferred from the stellar surface mass density using the ``extended Schmidt Law" from 
\cite{Shi-11} plus an estimate of the mass-loss rate in wind.  

This behavior could reflect the dynamic response of a gas regulator to variations in the inflow rate.  By driving a gas-regulator system with a periodic cold gas inflow, 
\citetalias{Wang-19a} explored how the SFR in the system would respond.  A periodic SFR, the amplitude of which depended on the ratio of the driving frequency to the inverse gas depletion time-scale.   This could reproduce the observed relation between the gas depletion times (either $\tau_{\rm dep}$ or $\tau_{\rm dep,eff}$) and the observed dispersion of $\Sigma_{\rm SFR}$ from galaxy to galaxy. 
%SFR, as well as the observational features of the  profiles in the population of galaxies.  
In addition, 
%combining the empirical function of extended Schmidt law from \cite{Shi-11}, i.e. the star formation efficiency (the ratio of cold gas mass to SFR) is proportional to $\Sigma_*^{1/2}$, 
the dynamical gas-regulator model can qualitatively explain the observed dependence of the SFMS on 
stellar mass and stellar surface density \citep[][]{Wang-18b, Davies-19,Wang-19a}. 

In more detail, \citetalias{Wang-19a} showed that the temporal variation of SFR on long timescales 
($>\tau_{\rm dep,eff}$) should closely follow the variations in cold gas inflow, while on short timescales (much less than $\tau_{\rm dep,eff}$) the variation of SFR is damped out, with a reduced amplitude (and a small phase shift).    \citetalias{Wang-19a}  explored the transfer function between the amplitude of variations in the inflow and the amplitude of the resulting variations in the star-formation rate, as a function of the driving frequency, that would be expected in the gas regulator model.  Although this depends in detail on the waveform of the inflow (their Fig 11), they derived an analytic relation (their Equation 9) for the simplest case of linear sinusoidal variations. In principle, the power spectrum distribution (PSD) of the star-formation history of a galaxy described by such a regulator should be given by the PSD of the inflow history multiplied by this simple transfer function.

The transfer function in \citetalias{Wang-19a} was based purely on the gas-regulator model with constant star-formation efficiency (SFE) (although a time-varying SFE was also considered).  In reality, the SFR
may be more bursty on short timescales if feedback
processes were included. Indeed, in the FIRE simulation, 
\cite{Sparre-17} found that the FIRE galaxies exhibit order-of-magnitude SFR variations over time-scales 
of only a few Myr, due to the effects of stellar feedback.  

Therefore, it is very interesting to have observational information on the temporal evolution of the SFR to understand the movement of galaxies
on the SFMS, and to uncover what kinds of physical processes govern the variation of SFR on different timescales. 

Hydro-dynamical simulations provide, in principle, one approach to study the variability of SFR, since
accurate SFHs can be produced for simulated galaxies. By using the small scatter of the stellar-to-halo 
mass relation, \cite{Hahn-19} constrained the timescale of star formation variability assuming different 
correlation coefficient between star formation and host halo accretion histories for SF central galaxies. 
However, the models in \cite{Hahn-19} only consider the variation of star formation on SFMS on one particular
timescale, ignoring the presumable contribution on other timescales.  By directly measuring the 
movement of galaxies on the SFMS at different timescales in the EAGLE simulation, \cite{Matthee-19} found that 
the variation of SFR of galaxies on the SFMS originates, in that simulation, from a combination of fluctuations on short 
time-scales (in the range of 0.2-2 Gyr), likely associated with self-regulation of star formation,  
and a dominant contribution on long timescale ($\sim$10 Gyr), likely related to the halo formation times. 

Observationally, the accurate SFH of individual galaxies 
are far from well determined,
especially on short timescales, 
\citep[e.g.][]{Papovich-01, Shapley-01, Muzzin-09, Ocvirk-06, Gallazzi-09, Zibetti-09,
Conroy-13, Ge-18, Leja-19}. This restricts the investigation of the variability in SFR of 
individual galaxies.  The ratio of a short-timescale SFR to the SFR averaged on a longer timescale, also known as the ``burstiness'',  has been proposed to probe the recent 
change of SFR \citep[e.g.][]{Sullivan-00, Boselli-09, Wuyts-11, Guo-16, Sparre-17,
Broussard-19, Faisst-19}. Indeed, \cite{Broussard-19} found that the distribution of burstiness characterizes 
the variability of recent star formation for a galaxy population, rather than the average value 
of burstiness. 

Based on the sSFRs traced by different SFR indicators, such as H$\alpha$, UV+IR and 
$u$-band, which effectively average the SFR on different timescales, \cite{Caplar-19} tried to use the scatter in the sSFR of the SFMS to constrain the power spectral distribution of temporal variations in the sSFR.  However, the measurements of UV+IR and $u$-band luminosity are quite sensitive 
to the dust attenuation corrections. This means that the scatter of the SFMS traced by these
luminosities may be strongly affected by the uncertainties from dust attenuation correction or other systematic effects between indicators. 
In addition, the possibility that there may be some intrinsic scatter of the SFMS that is unrelated to temporal variations, e.g. that some galaxies may
systematically lie above (or below) the SFMS throughout their lifetime, is not
considered in their analysis.  

In the first paper of this series \citep[][hereafter \citetalias{Wang-19b}]{Wang-19b}, we developed a new parameter
to quantify the change of SFR for individual galaxies, \RSFRL. This is defined to be the ratio of the SFR averaged 
within the last 5 Myr to the SFR averaged within the last 800 Myr, which we denoted SFR79. We called this parameter 
``the star formation change parameter'' (or simply the change parameter), rather than the previously 
used burstiness, because the SFR can both increase and decrease with time.  We found that observationally the change parameter 
can be well determined with an uncertainty of about 0.07 dex from the H$\alpha$ emission and H$\delta_{\rm A}$ 
absorption equivalent widths, plus the amplitude of the 4000 \AA\ break.  
This is because the H$\alpha$ emission line is a good tracer for the SFR of 
most recent 5 Myr, while H$\delta_{\rm A}$ absorption traces the SFH within the last $\sim$1 Gyr. 
In addition, since these three observational parameters are all relative quantities, and measured at 
essentially the same wavelength, they are insensitive to the dust attenuation, except to the extent that stellar populations of different ages experience different extinction, as discussed in \citetalias{Wang-19b}.

In \citetalias{Wang-19b}, we established an estimator of the \RSFRL\ based on the above three observational 
parameters, and applied it to the SF galaxies from MaNGA \citep[][]{Bundy-15}. 
Consistent with, but quite independent of, our analysis in \citetalias{Wang-19a}, we found in \citetalias{Wang-19b}  
that regions with shorter implied gas depletion timescales (again inferred from the extended Schmidt Law from the stellar surface mass density) show a larger amplitude of the change in SFR, i.e. a larger dispersion in SFR79.  In other words, we were able to confirm that the effects in \citetalias{Wang-19a}, from galaxy-to-galaxy, that we had {\it interpreted} as being due to temporal variations in SFR in individual galaxies, really were due to real temporal variations traced by \RSFRL.
This further supports the interpretation that the origin of the variation of SFR within and across galaxies is the dynamical response of gas regulator model to a time-varying
inflow, and further strengthens  
the notion that the narrow SFMS is indeed the result of the 
quasi-steady state between the inflow, outflow and star formation \cite[e.g.][]{Lilly-13, Wang-19a, Wang-19b}. 

In the current work, we will investigate the constraints that can be set on the power spectrum distribution (PSD) of the temporal variation of 
SFR in galaxies on the SFMS.  The (evolving) position of a galaxy relative to the SFMS is given by $\Delta$sSFR(t), which measures the log sSFR relative to the median sSFR of the SFMS at that stellar mass.
This is determined using the data from \citetalias{Wang-19b}. We will discuss the meaning of the PSD of $\Delta$sSFR(t).  For each individual SF galaxy, we have 
the measurements of SFR$_{\rm 5Myr}$, SFR$_{\rm 800Myr}$ and SFR$_{\rm 5Myr}$/SFR$_{\rm 800Myr}$.  
%The scatter of \RSFRL, SFR$_{\rm 5Myr}$ and SFR$_{\rm 800Myr}$ 
These contain the 
information about the temporal variations in the SFH on timescales greater than 5 Myr.  We can use this information on the temporal variation within individual galaxies to supplement that of variations (at a single epoch) across the galaxy population that comes from the scatter of sSFR on the SFMS.  
Interestingly, it is clear that the scatter of \RSFRL\  is unlikely to be affected by the sort of ``intrinsic scatter'' that may change the scatter of sSFR on the SFMS, discussed above. 
This enables us to consider a more detailed model than that in \cite{Caplar-19}.  
%In addition, the uncertainties of \RSFRL\ from both the observational measurements and the application of the estimator in \citetalias{Wang-19b} are much smaller than the observed scatter of \RSFRL, indicating the validity of our data. We refer the readers to \citetalias{Wang-19b} for more details. 

This paper is organized as follows. In Section \ref{sec:2}, we give a brief introduction 
to the data that are used. In Section \ref{sec:3}, we present the method to constrain the PSD of $\Delta$sSFR(t)
in this work.  In Section \ref{sec:4}, we explore our method by applying it first to the integrated 
quantities (measured within the effective radius) of the overall population of galaxies, lumping galaxies of all stellar masses together. 
In Section \ref{sec:5}, we present the main results of this work, 
by applying the method to the binned quantities within radial annuli 
derived in different stellar mass bins.  We discuss our result in Section \ref{sec:6} and summarize 
the work in Section \ref{sec:7}. 

For convenience, the average star-formation over the last 5 Myr, SFR$_{\rm 5Myr}$, is denoted 
as SFR7, and that averaged over the last 800 Myr, SFR$_{\rm 800Myr}$, as SFR9, 
and the ratio SFR$_{\rm 5Myr}$/SFR$_{\rm 800Myr}$ is denoted as SFR79. 
Following the \citetalias{Wang-19a} and \citetalias{Wang-19b}, 
we assume a flat cold dark matter
cosmology model with $\Omega_{\rm m}=$0.27, $\Omega_{\rm Lambda}$= 0.73 and $h$=0.7 
when computing distance-dependent parameters.  We will also assume the \cite{Chabrier-03} initial
mass function (IMF) and Cardelli-Clayton-Mathis dust attenuation curve \citep[CCM;][]{Cardelli-89}
throughout this work. As discussed in \citetalias{Wang-19b}, varying these assumptions produces a zero-point shift in SFR79 rather than altering the scatter in it, unless these quantities vary across the galaxy population. 

\section{Data} \label{sec:2}

\subsection{The galaxy sample} \label{subsec:2.1}

The sample of galaxies used in the present work is taken from \citetalias{Wang-19a} (also used in \citetalias{Wang-19b}),
and is based on the MaNGA spectra in the SDSS Data Release 14 \citep{Abolfathi-18}.  MaNGA is one of the
largest on-going integral field spectroscopic surveys, aimed at mapping $\sim$10,000 nearby galaxies with
two-dimensional spectra \citep{Bundy-15}. 
The wavelength covered by MaNGA is 3600-10300 \AA\ at R$\sim$2000, 
which enables the three diagnostic parameters, \dindex, \ewhda, and \ewhae, to be easily measured. 
Therefore we can apply the SFR79 estimator to the MaNGA data, and obtain the measurements of 
the SFR7 and SFR79, and therefore also SFR9, of galaxies. 

The galaxies are selected from the primary sample of the SDSS Data Release 14 of MaNGA, 
excluding galaxies that are mergers, irregulars, heavily disturbed 
galaxies \cite[see details in][]{Wang-18a}, as well as those with median S/Ns of 5500\AA\ 
continuum less than 3.0 at their effective radii. Quenched galaxies are also excluded based on 
the $M_*$($<$R$_{\rm e}$)-SFR($<$\re) diagram (see details in \citetalias{Wang-19a}), where the 
M$_*$($<$\re) and SFR($<$\re) are the stellar mass and SFR measured within effective radii.  
The final sample includes 976 SF galaxies, which is a representative sample of SFMS galaxies 
at low-redshift Universe.  We refer the readers to \citetalias{Wang-19a} and \citetalias{Wang-19b} for all
the details of the sample selection. 

\subsection{The measurements of SFR7 and SFR79} \label{subsec:2.2}

We presented the details of the measurements of SFR7 in \citetalias{Wang-19a}. These are calculated based on
the dust-corrected H$\alpha$ luminosity \citep{Kennicutt-98}.  The details of the more complex measurements of SFR79 
were also presented in \citetalias{Wang-19b} and will only briefly be summarized here. 
The SFR79 is based on three diagnostic observational parameters applied to the MaNGA data.  
We refer to the reader for section 2.3 in \citetalias{Wang-19a} and section 2 and 3 in 
\citetalias{Wang-19b} for all the details. 

We first constructed millions of mock SFHs of galaxies representing a 
huge range of the change parameter, SFR79. 
Then, we generated the dust-free spectra of these mock galaxies based 
on the single stellar population model using the code from \cite{Conroy-10}.  
We then measured these 
three diagnostic parameters based on the mock spectra and calculated the actual SFR79 from the mock SFHs. Finally we derived a 
solution for SFR79 from these three diagnostic parameters. 
This was done for a set of different stellar
metallicities.  When applying to the MaNGA data, we adopt an empirical stellar mass-metallicity 
relation from \cite{Zahid-17} to assign a stellar metallicity to each galaxy. 
The adopted estimators of SFR79 are presented in equation 2 and table 1 of \citetalias{Wang-19b}, 
with a typical uncertainty of 0.07 dex. 

In establishing the estimator, the three input diagnostic parameters are determined from dust free mock spectra, so the corresponding measurements on the observed spectra
should be corrected for dust attenuation, and specifically for differential, age-dependent, extinction.  We refer the readers to the details of this 
correction given in section 2.5 of \citetalias{Wang-19b}, which adopts a stellar age dependent dust model 
from \cite{Charlot-00}.  Here we do not repeat these details. 

When applying the estimator to the individual spaxels in MaNGA galaxies, we saw a weak
positive correlation between stellar surface density ($\Sigma_*$) and SFR79 (see figure 6 in \citetalias{Wang-19b})
, i.e. an overall negative radial gradient in SFR79.  
This negative radial gradient in SFR79 is not likely to be real, because a small gradient 
in SFR79 would result in a significant change in the overall radial gradient of 
SFR7 only assuming that the small gradients of SFR79 can 
be sustained for a few Gyrs.  The negative radial gradient in SFR79 may due to the fact 
that we do not consider any radial variation of stellar 
metallicity \citep[e.g.][]{Zheng-17, Goddard-17},  and a (possible) radial variation 
of the initial mass function \citep{Gunawardhana-11, Zhang-18}. Therefore we performed   
an {\it ad hoc} correction in SFR79 to eliminate this small dependence on $\Sigma_*$ (see details
in section 3.2 in \citetalias{Wang-19b}). Although this correction is arbitrary, we argued that this 
would not change any of the conclusions in \citetalias{Wang-19b}, and it also has a negligible effect 
on the PSD of specific SFHs of galaxies in this paper (as directly checked in Section \ref{subsec:4.2}). 

As discussed in \citetalias{Wang-19b},  we would expect more variation in SFR79 (or SFR7)  at 
spaxel-size scales due to the duty
cycle of star formation: in the star-formation duty cycle some small regions will be actively forming 
stars (with consequently high SFR79), while other regions are in a relative low SFR state, with a small negative SFR79.   Therefore, 
to reduce this {\it spatial} effect,  we performed the spatial binning scheme in \citetalias{Wang-19b}, and also in the present work. 

Here we briefly describe how we calculated the three diagnostic parameters in a given region,
and refer for more details to section 3.3 in \citetalias{Wang-19b}.
For \dindex,  we first calculated the flux density of the corresponding 
blue and red bandpass near 4000 \AA\ for each spaxel in a given region (radial bin). 
Then the \dindex\ of this region is simply the ratio of total flux density of red 
bandpass to the blue bandpass for all spaxels located in this bin.  The \ewhda\ and \ewhae\ are 
calculated in a similar way for a given region. 
In this process, we also estimated the purely observational errors of the three diagnostic parameters, 
and obtained the error of SFR79 by error propagation. In calculating the errors, we adopt the  
empirical function of \cite{Law-16} to correct the covariance of errors in the MaNGA spectra of
adjacent spaxels (see equation 6 in \citetalias{Wang-19b}). 

In \citetalias{Wang-19b}, we performed two types of binning schemes: binning all the spaxels within \re\ as a 
global measurement of SFR79 in the galaxy, and binning those within radial annuli of 
width 0.2\re.  In the present work, we focus on the integrated
measurements within \re\ and within the five radial annuli within \re. This is 
because 1) almost all the galaxies are covered within \re\ in the observations, 2) the 
stellar metallicity of the outer regions may be significantly lower than the assigned ones in computing the SFR79, 
and 3) the star formation of the outer regions are more likely to be affected by environmental effects,
which are not considered in the present work. 

\subsection{The diagram of $\Delta$sSFR7 vs. $\Delta$sSFR9} \label{subsec:2.3}

\begin{figure}
  \centerline{
    \includegraphics[width=0.42\textwidth]{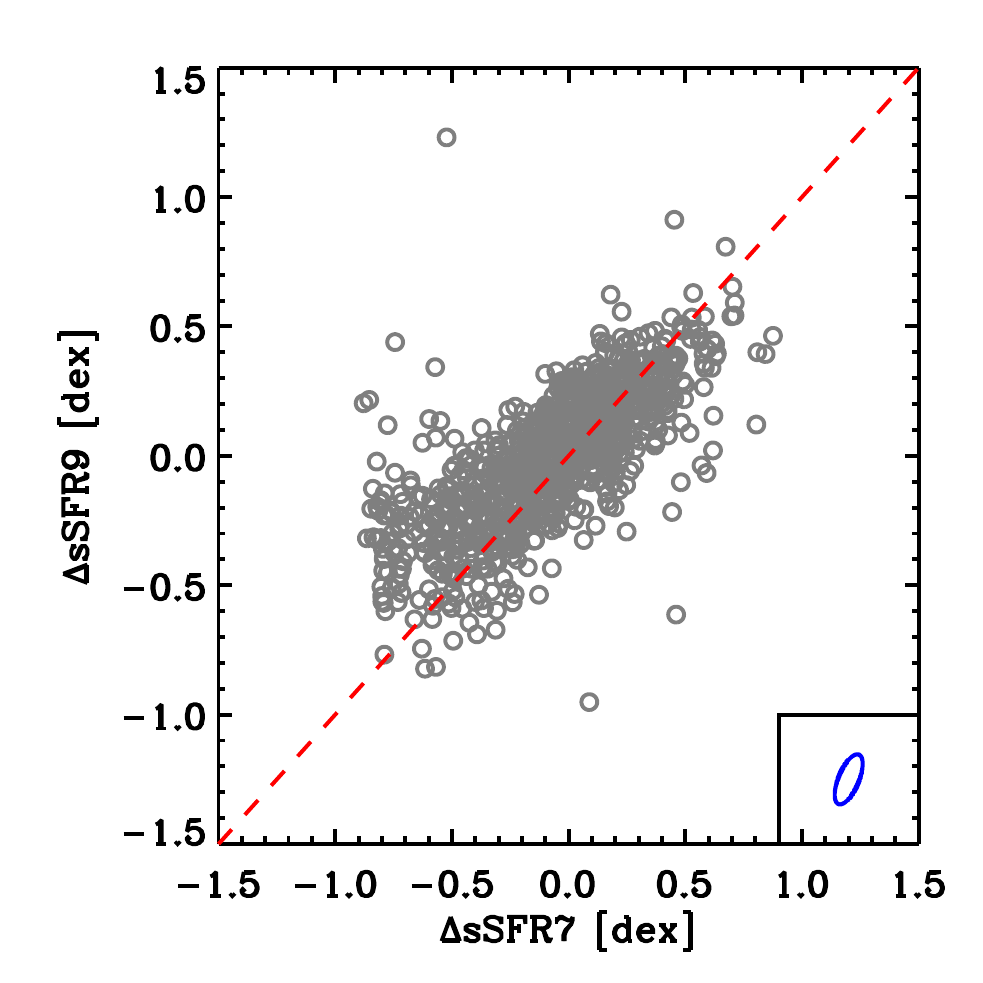}
  }
  \caption{ The $\Delta$sSFR7 versus $\Delta$sSFR9 for the sample galaxies, taken
  from the figure 10 of the \citetalias{Wang-19b}.  Each galaxy in the sample is represented by a single dot based on its integrated spectrum within \re.
  The median value of  $\Delta$sSFR7 and $\Delta$sSFR9 is shifted to be zero. 
  The blue ellipse shows the typical uncertainties of the data points, 
  and combines the uncertainties from the input observational measurements and from the calibration of SFR79. }
  \label{fig:global}
\end{figure}

\begin{figure*}
  \begin{center}
    \includegraphics[width=0.99\textwidth]{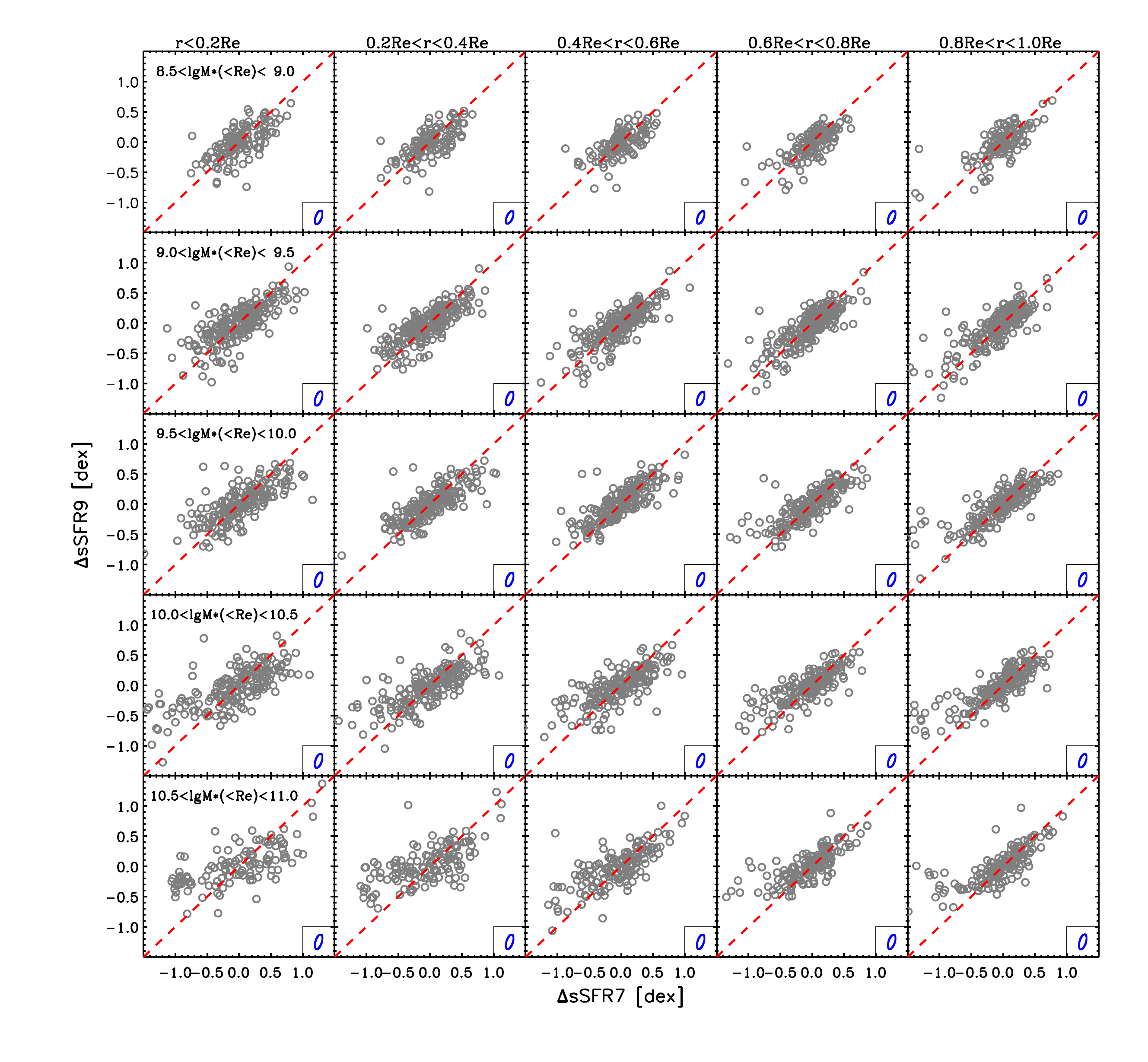} 
    \end{center}
  \caption{As for Figure \ref{fig:global} but now the sample of MaNGA galaxies is split into five stellar mass bins (top to bottom) and the $\Delta$sSFR7 and $\Delta$sSFR9 are calculated for each galaxy from the integrated spectra within five annular regions of width 0.2 \re\ (increasing radius from left to right). 
  %As in Figure \ref{fig:global}, in each panel, the median value of $\Delta{\rm sSFR7}$ 
  %and $\Delta{\rm sSFR9}$ are shifted to be zero.  In the bottom right of each panel,
  %the ellipse indicates the typical uncertainty of the data points, including the uncertainties from the observations and from the estimator of %SFR79. 
  }
  \label{fig:data_resolved}
\end{figure*}

Based on the SFR7 and SFR79 measured within \re, we can then obtain the SFR9 of each individual 
galaxy. These can then be used, together with the stellar masses measured within \re, to compute specific SFR, and to define the overall SFMS of the population on both timescales. We then define  $\Delta$sSFR7 (and $\Delta$sSFR9) to be the vertical deviation (i.e. in sSFR) of a galaxy from the 
median SFR7-based (or SFR9-based) SFMS.   

Figure \ref{fig:global} shows the correlation between $\Delta$sSFR7 and $\Delta$sSFR9 (measured within \re) for our
sample galaxies, taken from figure 10 of \citetalias{Wang-19b}. 
In the bottom right corner, we indicates the median uncertainty of the data 
points, including both observational uncertainties in the input diagnostic parameters and the uncertainty from the estimator itself, as 
in \citetalias{Wang-19b}.  Specifically, the typical uncertainty of SFR7 is 0.06 dex, which is due to the
variations in electron temperature in the range of $T_{\rm e}$=5000-2000 K \citep{Osterbrock-gn}
when converting the H$\alpha$ luminosity to SFR7 using the empirical relation from \cite{Kennicutt-98}.
The median uncertainty of SFR79 due to observational uncertainties is 0.042 dex, and 
due to the estimator itself is 0.063 dex. 
Therefore, the overall typical uncertainty 
in SFR79 is $\sim$0.076 dex.  

In Section \ref{sec:4}, we will constrain the PSD of $\Delta$sSFR(t) by matching the 
galaxy distribution on the $\Delta$sSFR7-$\Delta$sSFR9 diagram in Figure \ref{fig:global}, taking  
the uncertainties in the data points into account, by convolving the predicted galaxy distribution on
the $\Delta$sSFR7-$\Delta$sSFR9 diagram with the uncertainty 
ellipse denoted in the bottom right corner of Figure \ref{fig:global}. 

In principle, Figure \ref{fig:global} contains the key information needed to constrain the PSD of the sSFHs. 
Four inter-related quantities related to the variability of sSFH can be read from Figure \ref{fig:global}: 
(i) the overall scatter of $\Delta$sSFR7 ($\sigma_{\rm 7}$), i.e. the scatter of the SFMS on this timescale, 
(ii) the equivalent overall scatter of $\Delta$sSFR9 ($\sigma_{\rm 9}$), 
(iii) the scatter of SFR79 ($\sigma_{\rm 79}$), i.e. the scatter perpendicular to the dashed line, and 
(iv) the covariance 
of $\Delta$sSFR7 and $\Delta$sSFR9. These four parameters are not independent, 
and their inter-relation can be written as: 
\begin{equation} \label{eq:1}
\sigma_{79}^2 =  \sigma_7^2 + \sigma_9^2 - 2{\rm Cov}(\Delta {\rm sSFR7}, \Delta {\rm sSFR9}) 
\end{equation}
where the Cov($\Delta$sSFR7, $\Delta$sSFR9) is the covariance of $\Delta$sSFR7 and $\Delta$sSFR9. 
This appears to result in three independent parameters related to the variability of sSFH. 
However, as we will discuss later in Section \ref{subsec:3.2}, the $\sigma_7$, $\sigma_9$
and $\sigma_{79}$ are still not fully independent. 

We comment here on a small but significant detail.  The $\sigma_7$ (or $\sigma_9$) is taken to be the scatter of $\Delta$sSFR7 (or $\Delta$sSFR9). This 
quantity is calculated, in practice, from the average star-formation rate within the most recent 5 Myr (or 800 Myr)
divided by the {\it current} stellar mass.  It is therefore not precisely the same as what would be ideally desired, which would be the average sSFR, averaged over the last 5 Myr (or 800 Myr), simply because the stellar mass will have slightly increased during this time interval, especially for the longer averaging timescales.  
However, given the current relatively low sSFR$\ll$1 Gyr$^{-1}$ of the SFMS galaxies, the difference due to the mass increase is anyway small, less than a few percent.  
Furthermore, this need not be a concern in practice because, in modelling the distribution of data points on the 
$\Delta$sSFR7-$\Delta$sSFR9 diagram, we are anyway free to use exactly the same approach 
to obtain the $\Delta$sSFR7 and $\Delta$sSFR9 as that in the observations, i.e. we can choose to divide the average star-formation rate by the current mass 
(see details in Section \ref{sec:3}). We will discuss this further in Section \ref{subsec:3.1}. 

In \citetalias{Wang-19a} and \citetalias{Wang-19b}, we presented evidence that the star formation at different 
galactic radii of galaxies was varying up and down relative to the median $\Sigma_{\rm SFR}$ profile, with an amplitude that appeared to strongly increase with decreasing implied (effective) 
gas depletion time.  Therefore, it will be of interest later in the paper to constrain the PSD of sSFH at 
different galactic radii in order to explore whether there is 
any correlation between the PSD and the gas depletion time.  Accordingly, we will calculate
the sSFR7 and sSFR9 at five different galactic radial bins for each galaxy, and compare these quantities for all the galaxies in five stellar mass bins.  

Figure \ref{fig:data_resolved} shows the $\Delta$sSFR7-$\Delta$sSFR9 diagram for these separate regions, separated by the overall (within \re) stellar
mass of the galaxy and by the relative radial location. 
For each radial bin of each individual galaxy,  we define the $\Delta$sSFR7 (or $\Delta$sSFR9) 
as the deviation from the median sSFR7 (or median sSFR9) in the same radial bin of 
the galaxy population in the corresponding stellar mass bin.  Therefore, the $\Delta$sSFR7 (or
$\Delta$sSFR9) have always, by definition, a median value of zero. 
The plots are displayed with increasing radius from left to right, and with increasing stellar 
mass from top to bottom. 
In the similar way of Figure \ref{fig:global}, we indicate the typical uncertainty of the data points 
in each panel, including both the error from observations and from the
the estimator itself. 

In Section \ref{sec:5}, the PSD of $\Delta$sSFR(t) for each galactic radius for each overall stellar mass bin 
will be constrained by matching the distribution of data points on the corresponding
$\Delta$sSFR7-$\Delta$sSFR9 diagram in the respective panels in Figure \ref{fig:data_resolved}.  Comparing these 25 panels, it is immediately apparent that the distribution of points is different in the different panels, and we would expect the PSD to therefore be subtly dependent on both the overall mass of the galaxy and the radial location of the regions.

\section{Constraining the PSD} \label{sec:3}
\label{}

\subsection{The concept of the PSD} \label{subsec:3.1}

For continuous signals over all time, like stationary processes\footnote{A stationary process is a 
stochastic process whose unconditional joint probability distribution does not change when shifted in time.
Therefore, in the stationary process, parameters such as the mean and variance also do not change over time.}, 
the power spectrum distribution characterizes 
how the power of a signal or time series is distributed in frequency.  

In the present work, we aim at investigating 
the PSD of the deviations of log sSFR(t) from the average SFMS at that mass, which we have denoted $\Delta$sSFR(t).  We can do this for either the overall galaxy population or of particular subsets of it chosen by overall mass and relative radius.  
We focus on the PSD of the sSFR {\it relative} to the average SFMS, rather than the PSD of the SFR(t) or sSFR(t) directly,  
because the scatter of SFMS is observed to be nearly independent of time \citep{Kelson-14, Kelson-16, Caplar-19, Hahn-19}, whereas both the mean sSFR and mean SFR do change with time. This suggests that it is the variations of $\Delta$sSFR(t) that can best be characterized as a stationary process. For brevity, the $\Delta$sSFR(t) will also be denoted as $y(t)$ in this work. 

Assuming an individual SF galaxy with a known variation relative to the average (median) SFMS, $y(t)$ (= $\Delta$sSFR(t)), 
the truncated Fourier transform $\hat{y}(\nu)$, in which the signal is integrated only over a finite interval of time [0, T], can be written as:
\begin{equation} \label{eq:2}
    \hat{y}(\nu) =  \frac{1}{\sqrt{T}}\int_{0}^{T} {y}(t) e^{-i2\pi\nu t} dt. 
\end{equation}
Then by definition, the PSD of $y(t)$ as a function of frequency can be written as: 
\begin{equation} \label{eq:3}
\begin{split}
{\rm PSD}(\nu) &= \lim_{T\to\infty} \textbf{E}[|\hat{\rm y}(\nu)|^2] \\
               &= \lim_{T\to\infty} \frac{1}{T} \int_{0}^{T} \int_{0}^{T} 
                  \textbf{E}[y(t)y(t')]e^{-i2\pi\nu(t-t')} dt dt' 
\end{split}
\end{equation}
where the \textbf{E} denotes the expected value.  
For a stationary process, one can make the change of variables 
$\Delta$t=t$-$t$'$, and then it is evident that the PSD of $y(t)$ and the autocorrelation
function of $y(t)$ are a Fourier-transform pair, as in the Wiener-Khinchin theorem. 

Therefore, according to Equation \ref{eq:3},  the ratio of the SFR averaged within the most recent 5 Myr 
to the SFR averaged within that same 5 Myr interval at a given lookback time, e.g. $\sim$1 Gyr previously, 
is more directly relevant for the PSD than the empirically defined change 
parameter, \RSFRL, that we introduced in \citetalias{Wang-19b}. 
%This compares the SFR averaged within the most recent 5 Myr to that averaged over the $\sim$1 Gyr timescale.  
Ideally, we would want to measure the SFR on 5 Myr timescales 1 Gyr ago but,  observationally, this is not currently possible, and it will likely be very challenging for a long time to come.

The power in $y(t)$ within a given frequency band [$\nu_1$, $\nu_2$], can be written as: 
\begin{equation} \label{eq:4}
   P_{[\nu_1,\nu_2]} =  2\int_{\nu_1}^{\nu_2} {\rm PSD(\nu)}d\nu. 
\end{equation}
The factor of two comes from an equal amount of power that can be attributed to positive and 
negative frequencies. The power basically characterizes the contributions to the variation 
of $y(t)$ in the given frequency band, or equivalent given timescale band. 
The main goal of this work is to constrain the PSD($\nu$) of $\Delta$sSFR(t), which will tell 
us the contributions to the variation of $\Delta$sSFR(t) from different timescales (or frequencies). 
However, we note that the power defined in Equation \ref{eq:4}, when integrated from zero to 
1/5 Myr$^{-1}$ (or 1/800 Myr$^{-1}$), is close but not exactly equal 
to the variance of $\Delta$sSFR(t) averaged with 5 Myr (or 800Myr), because the Fourier 
transform of a step function in time is not exactly a step function in frequency. 
In practice, when performing the integration in the right side of Equation \ref{eq:4}, 
the lower boundary of the integration could not reach zero, because the lifetime of galaxies 
should be less than the age of Universe. Therefore, frequencies lower than the inverse of 
the Hubble time ($\tau_{\rm Hubble}$) are meaningless.  
We will come back to this point in Section \ref{subsec:3.2}.  

Since we plan to constrain the variations of galaxies on the SFMS, one might imagine that the overall stellar mass 
of the galaxies could be another parameter to constrain the variability of $\Delta$sSFR(t) on very long timescales, 
since the mass reflects the average SFR when averaged on a timescale of $\sim$10 Gyr. 
Following the definition of SFR7 and SFR9 in Section \ref{subsec:2.2}, 
we could then define the SFR10 to be the SFR averaged within the Hubble time divided by the current mass.  
It is then however evident that the value of sSFR10 will be identically equal to the inverse age of the universe, $\tau_{\rm Hubble}^{-1}$, and therefore the same for all galaxies. 
The scatter of $\Delta$sSFR10, which we could notate as $\sigma_{10}$, is therefore identically zero.   It is therefore not possible to use the stellar mass to 
constrain the variability of $\Delta$sSFR(t) on very long timescales.  

We note that although $\sigma_{10}$ (as defined here) is identically zero, there is no requirement that 
%the dispersion 
all galaxies have the same average $\Delta$sSFR(t) when averaged over the Hubble time.  Quite the contrary, 
we could well imagine that some galaxies could lie above or below the SFMS throughout their lifetimes.  This introduces the important concept of ``intrinsic'' scatter of the SFMS, i.e. a dispersion in the sSFR of the population (at given mass) that is {\it unrelated} to any temporal variability of individual objects. This will
be discussed further in Section \ref{subsec:3.2}.  

\subsection{Ergodicity}  \label{subsec:3.2}

In this work, we only have the SFR7, SFR9, and SFR79 for the sample galaxies 
at a single epoch, i.e. the current epoch, rather than the full time history of star-formation that would be obtained by
monitoring an individual galaxy for a very long time\footnote{This is not possible in 
the observations, but is in principle achievable in simulations.}.  This introduces the question of the
ergodicity of the chosen galaxy population when attempting to constrain the
PSD of individual sSFH from the snapshot of the population of galaxies at a single epoch. 

In probability theory, an ergodic dynamical 
system has the same behavior averaged over time as averaged over all the possible
states at a given time.  The assumption of ergodicity requires that
the distribution of $y(t)$ for an individual galaxy in time is the same
as the distribution of $y(t_{\rm 0})$ 
across the galaxy population at a given fixed time $t_{\rm 0}$.  If this is satisfied, then studying the variation of $y(t_{\rm 0})$ over
the galaxy population at fixed epoch is equivalent to monitoring an individual 
galaxy over a very long time. 

Can we assume that SFMS galaxies are ergodic, with a $\Delta$sSFR(t) constructed from a single PSD, 
and that there are no other (non-temporal) sources of variation in their sSFR? If yes, we can directly link the $\sigma_{7}$, and $\sigma_{9}$ with 
the PSD.  The $\sigma_{7}$ and $\sigma_{9}$ can be expressed as follows: 
\begin{equation} \label{eq:5}
\begin{split}
    \sigma_{7}^2 & \sim  2\int_{\tau_{\rm Hubble}^{-1}}^{1/5 {\rm Myr}^{-1}} {\rm PSD(\nu)}d\nu  \\
    \sigma_{9}^2 & \sim  2\int_{\tau_{\rm Hubble}^{-1}}^{1/800 {\rm Myr}^{-1}} {\rm PSD(\nu)}d\nu
\end{split}
\end{equation}
The left part and the right part of Equation \ref{eq:5} are not exactly the same, 
due to the facts that were mentioned in Section \ref{subsec:2.3} and \ref{subsec:3.1}:
1) the Fourier transform of a step function in time is not exactly a step function 
in frequency, and 2) the $\Delta$sSFR7 and $\Delta$sSFR9 (as defined and operationally determined in terms of the {\it current} stellar mass) are not identical to the average $\Delta$sSFR averaged over the last 5 Myr and 800 Myr respectively. 
Again, we note that this is not a concern when constraining the PSD, because we 
can include these effects when matching the predicted distributions of observable quantities with the observations. 

Furthermore, in the real Universe, it may well be that galaxies are not ergodic.
Galaxies of different stellar mass, structure, or environment, may have different mean $\Delta$sSFR and different
PSDs describing their time variability about this mean. 
For instance, as already noted, some galaxies may have 
a systematically higher sSFR(t) (or lower sSFR(t)) than the average SFMS throughout cosmic time, perhaps because they reside in particular environments.  This
would result in an additional component to the scatter of the SFMS at a given epoch that was wholly unrelated to the time variations of individual galaxies represented by the PSD.

Similar effects could in practice be introduced by the observational methodology itself: at the very least, any significant observational error in the determination of the (current) stellar masses of individual galaxies, even if completely random, will introduce a scatter into the SFMS, that may be completely unrelated to any time variation in the SFR (or even worse, might be related in a highly complicated way).   Any error in the estimation of stellar mass of a galaxy would change both $\Delta$sSFR7 and $\Delta$sSFR9 (by the same amount), but not the SFR79 introduced in \citetalias{Wang-19b}.   

More subtle violations are also possible. The simple fact that the distributions of points on the different panels of $\Delta$sSFR7-$\Delta$sSFR9 on Figure \ref{fig:data_resolved} are slightly different shows that an assumption of ergdocity across the whole galaxy population cannot be valid, since it is clear that different mass galaxies must explore slightly different areas of parameter space.  Lumping them all together, as on Figure \ref{fig:global} and deriving a single PSD from the combined distribution, would clearly not be strictly correct.

How can we best deal with the unknown degree of violation of ergodicity that will be present in any chosen population of galaxies, e.g. within Figure \ref{fig:global} or within each panel of Figure \ref{fig:data_resolved}, from whatever source?    In this work, we will try to 
take this non-ergodicity into consideration by introducing an additional parameter, which we call the
``intrinsic scatter" of the SFMS, $\sigma_{\rm int}$.  We imagine that this represents all those contributions to the variation of the $\Delta$sSFR within the population at a given epoch that are unrelated to the temporal variations of individual galaxies that we assume may be represented by a single PSD.  

Due to the plausible existence of a $\sigma_{\rm int}$ term, it is dangerous to simply use the scatter of the SFMS, as indicated by different SFR indicators that average on different timescales, to constrain the time variability of the SFH \citep[c.f.][]{Caplar-19}.   This is quite separate from the additional worry that the SFMS scatter based on these different SFR indicators may be contaminated to varying degrees by uncertainties in the dust attenuation corrections, or by variations in the SFR calibrators used for each. 

We noted above that $\sigma_{\rm int}$ contributes to the 
variance of $\Delta$sSFR7 and $\Delta$sSFR9, in the sense that, the right part of Equation \ref{eq:5}
should include the $\sigma_{\rm int}^2$.   But, with our definition, $\sigma_{\rm int}^2$ does not contribute to the variance of SFR79.  Given also that the star formation change parameter SFR79, defined and calibrated in \citetalias{Wang-19b}, is much less sensitive to dust attenuation than e.g. ultraviolet luminosities, this is a further reason why the SFR79 information
developed in \citetalias{Wang-19b} is a very important tool to study the variability of sSFH \citep[c.f.][]{Caplar-19}. 

In the remainder of this work, we will not consider other possibilities for non-ergodicity.   We will however return to this issue again in Section \ref{subsec:6.1}.  
We note here that the addition of the intrinsic scatter in this way is, in principle, equivalent to adding Fourier modes of a PSD at very low frequencies, i.e. much lower than the inverse age of the Universe ($\nu<\tau_{\rm Hubble}^{-1}$). 

In \citetalias{Wang-19b}, we showed that the SFR79 does not correlate with the $\Delta$sSFR9, i.e. the 
{\it change} of star formation does not depend on the average position (within the last 800 Myr) 
of the galaxy on the SFMS.  This is required if the scatter of the SFMS is to stay more or less constant with time, 
since a positive (or negative) correlation between SFR79 and $\Delta$sSFR9 would 
result in a broadening (or narrowing) of the SFMS. As we mentioned in Section \ref{subsec:2.1}, 
the constancy of the dispersion of the SFMS indicated that $\Delta$sSFR(t) be described as a stationary process. This puts 
extra-constraints on $\sigma_7$, $\sigma_9$, and $\sigma_{\rm 79}$, which makes them to be not
fully independent. This further limits the number of freedom in forms for the PSD that may realistically be considered in the present work. 

\subsection{Possible forms of the PSD}

\label{subsec:3.3.0}

In principle, the form of the PSD for an unknown time-series process can be arbitrary. However, a simple power-law form of the PSD, 
PSD $\propto \nu^{-\alpha}$, has often used to characterizes stochastic processes. The slope $\alpha$ determines the relative contribution of variations at short and long timescales. White noise has $\alpha=0$ and is a perfectly uncorrelated time-series process at any given time separation. 
The cases of $\alpha=1$ and $\alpha=2$ are sometimes called pink noise and red noise respectively. The latter is also known as a random walk or Brownian motion.

\cite{Caplar-19} adopted a broken power-law to characterize the PSD of the variation of galaxies on the SFMS, i.e. PSD $= \sigma^2/(1+(\tau_{\rm break}\nu)^\alpha)$.  The $\sigma$ defines the overall amplitude of the PSD, and $\tau_{\rm break}$ characterizes the timescale beyond which the process becomes uncorrelated.  At the $t\gg\tau_{\rm break}$ (or $\nu\ll \tau_{\rm break}^{-1}$), the PSD approximates to a constant value, while at  $t\ll\tau_{\rm break}$ (or $\nu\gg \tau_{\rm break}^{-1}$), the PSD approximates to a single power function with a slope of $\alpha$.  
A broken-power law PSD can greatly reduce the power at low frequencies, leading to convergence of the integral of the PSD at the low frequency end.  

In principle, we could also adopt the same form of PSD in this work, and treat the power in the PSD at $\nu<\tau_{\rm Hubble}$ as the contribution to the ``intrinsic scatter'' discussed above. 
As noted above, any power on timescales greater than $\tau_{\rm Hubble}$ has a similar effect as the intrinsic scatter mathematically, although it is meaningless physically. 

However, treating the effects of $\sigma_{\rm int}$ by unphysical low frequency modes seems to us dangerous, since it effectively requires that these unphysical low frequency modes, which are required to mimic any intrinsic scatter, be continuous in power with the physically meaningful modes on timescales below the age of the universe.  
Similarly, using
a broken power-law that extends into unphysically low frequencies, may in practice return break frequencies that have more to do with mimicking the effects of any intrinsic scatter than indicating the timescale of any real physical process.

Trying to include very low frequency modes in the PSD, one furthermore encounters the practical difficulty of needing to generate an infinitely long time-series in the realization of mock galaxies. 

Therefore, in this work, we adopt for simplicity a single power PSD of $\Delta$sSFR(t), truncated to zero at all timescales greater than $\tau_{\rm Hubble}$.  We then allow an extra-parameter, the ``intrinsic scatter'' of the SFMS, whose amplitude is allowed to vary completely {\it independently} of the PSD.

%The broken power-law, the predication in W19 by the gas regulator model.  We find that ...
%While the broken time is 2$\pi\tau_{dep}$, which is comparable to the age of universe. 

\subsection{Illustration with a single power-low PSD}  \label{subsec:3.3}

\begin{figure*}
  \begin{center}
    \includegraphics[width=0.95\textwidth]{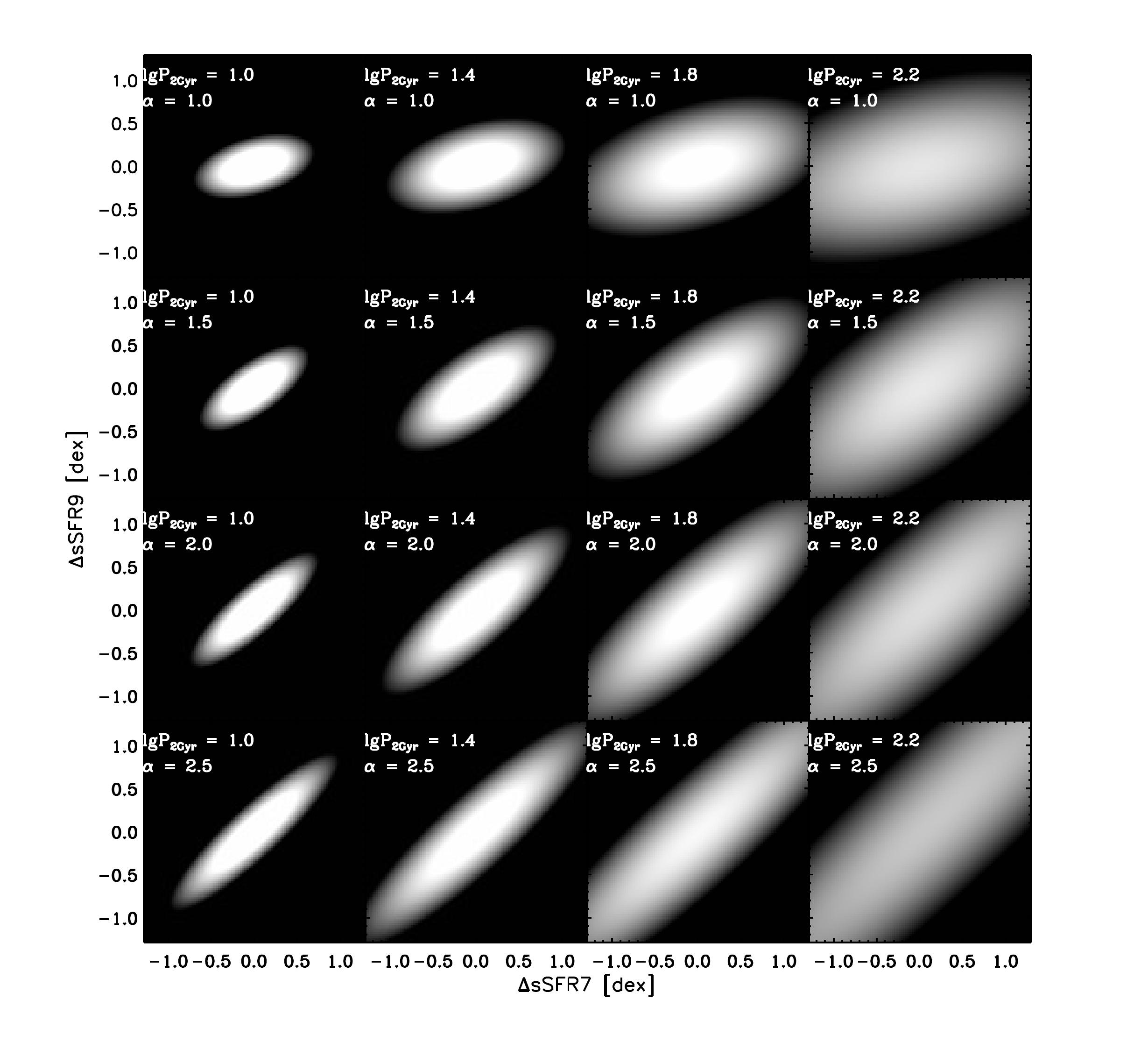} 
    \end{center}
  \caption{The probability distribution of mock galaxies on the 
  $\Delta$sSFR7-$\Delta$sSFR9 diagram for different input PSDs.
  The plots are arranged with increasing amplitude of the PSD, $P_{\rm 2Gyr}$, 
  from left to right, and with increasing slope of the PSD, $\alpha$, from 
  top to bottom. In each panel, the values of $\alpha$ and $P_{\rm 2Gyr}$ are denoted 
  in the top left corner. These  distributions of mock galaxies have not been convolved with observational uncertainties, but these are 
  included when these distributions are compared with the observational data from Figures \ref{fig:global} and \ref{fig:data_resolved}, in Section \ref{sec:4} and \ref{sec:5} of the paper.}
  \label{fig:example}
\end{figure*}

\begin{figure*}
  \begin{center}
    \includegraphics[width=0.45\textwidth]{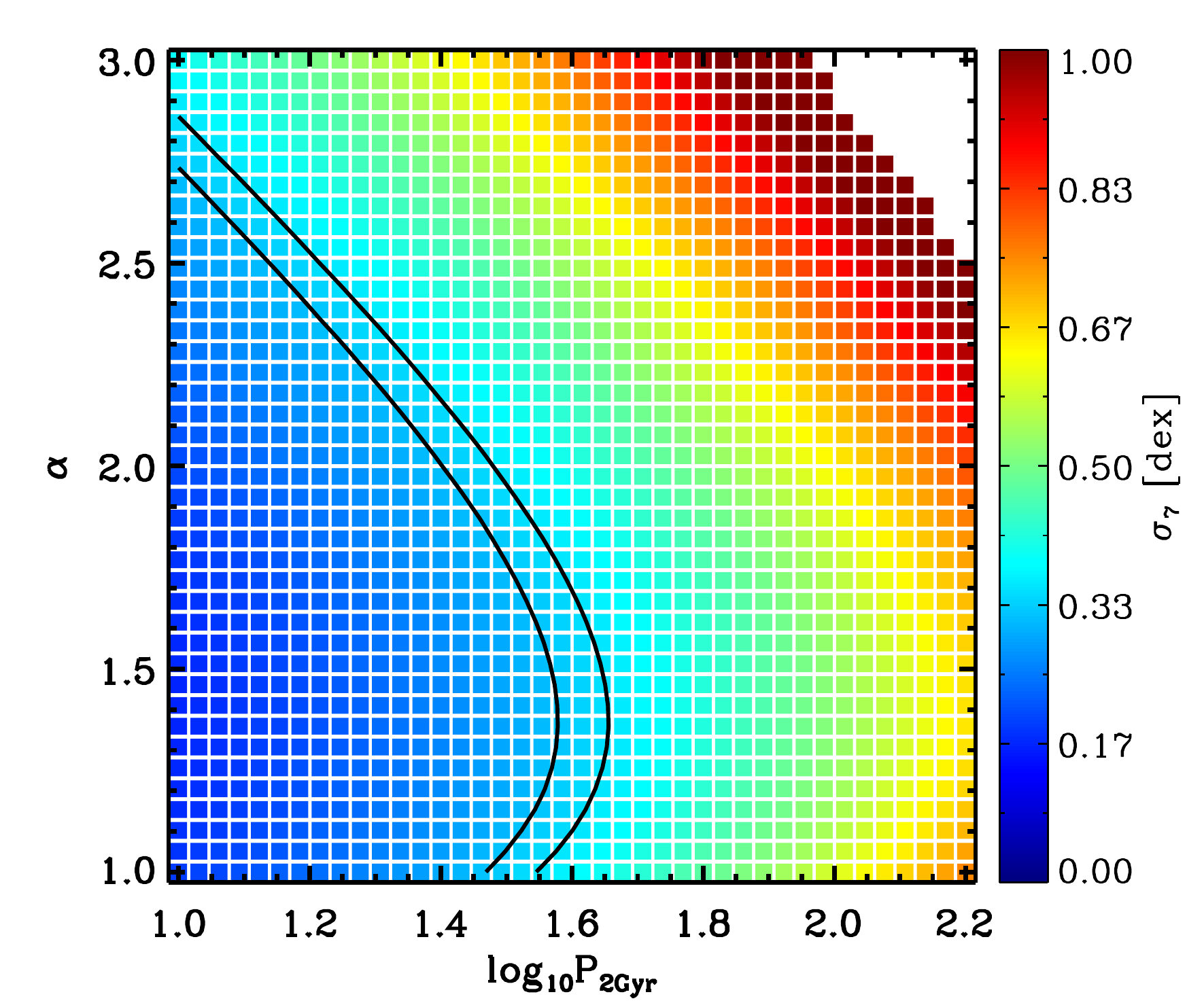}
    \includegraphics[width=0.45\textwidth]{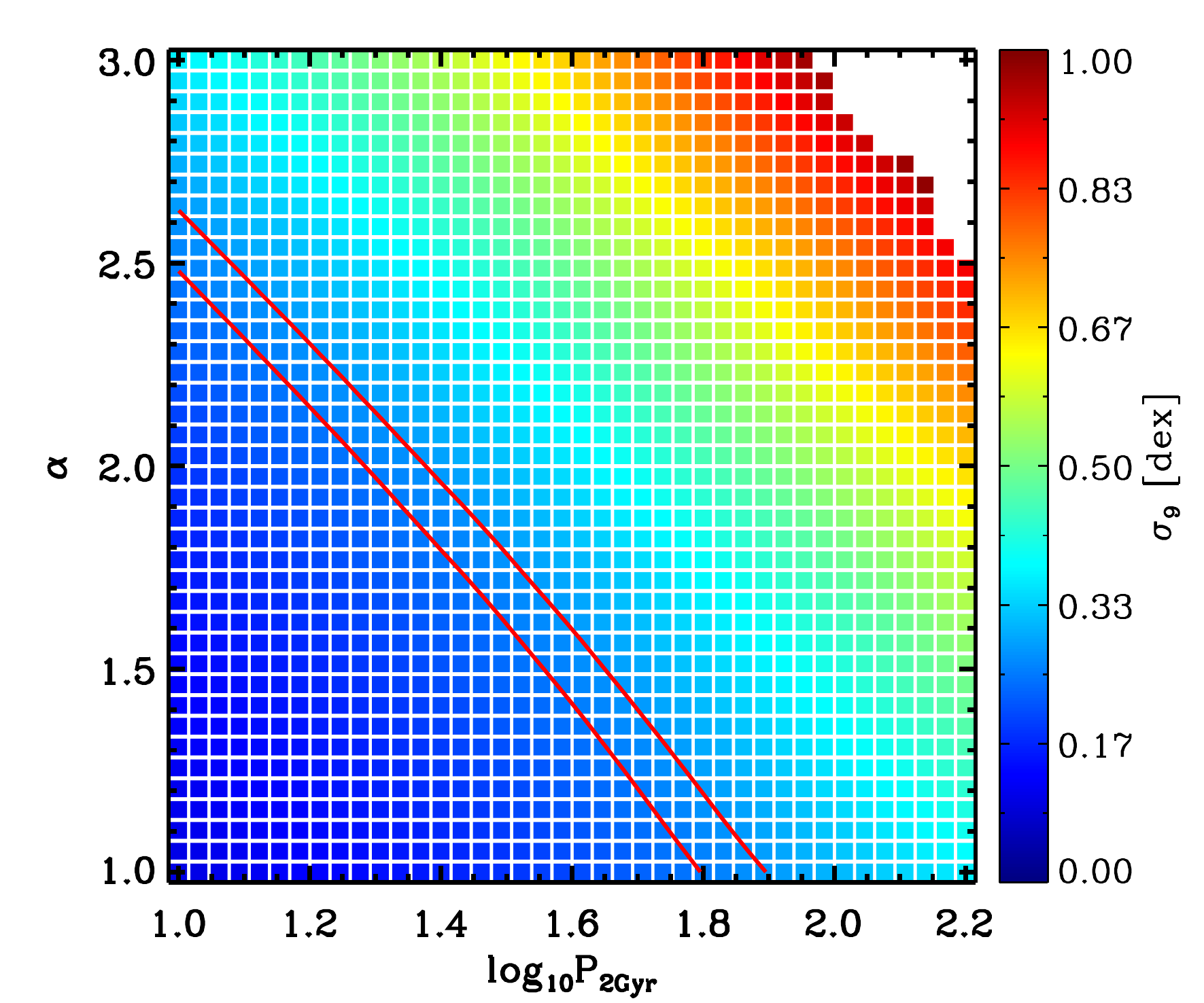}
    \includegraphics[width=0.45\textwidth]{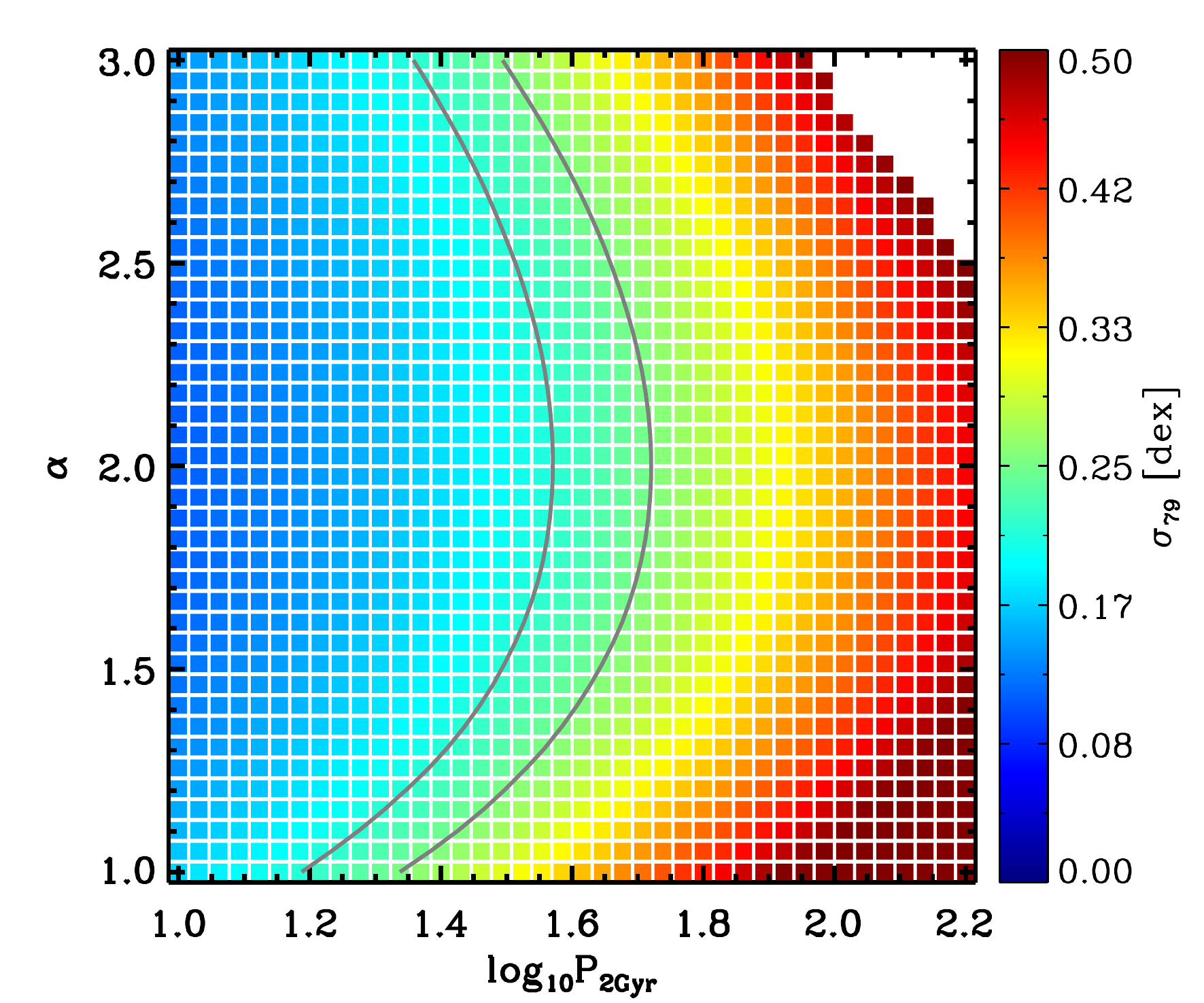}
    \includegraphics[width=0.45\textwidth]{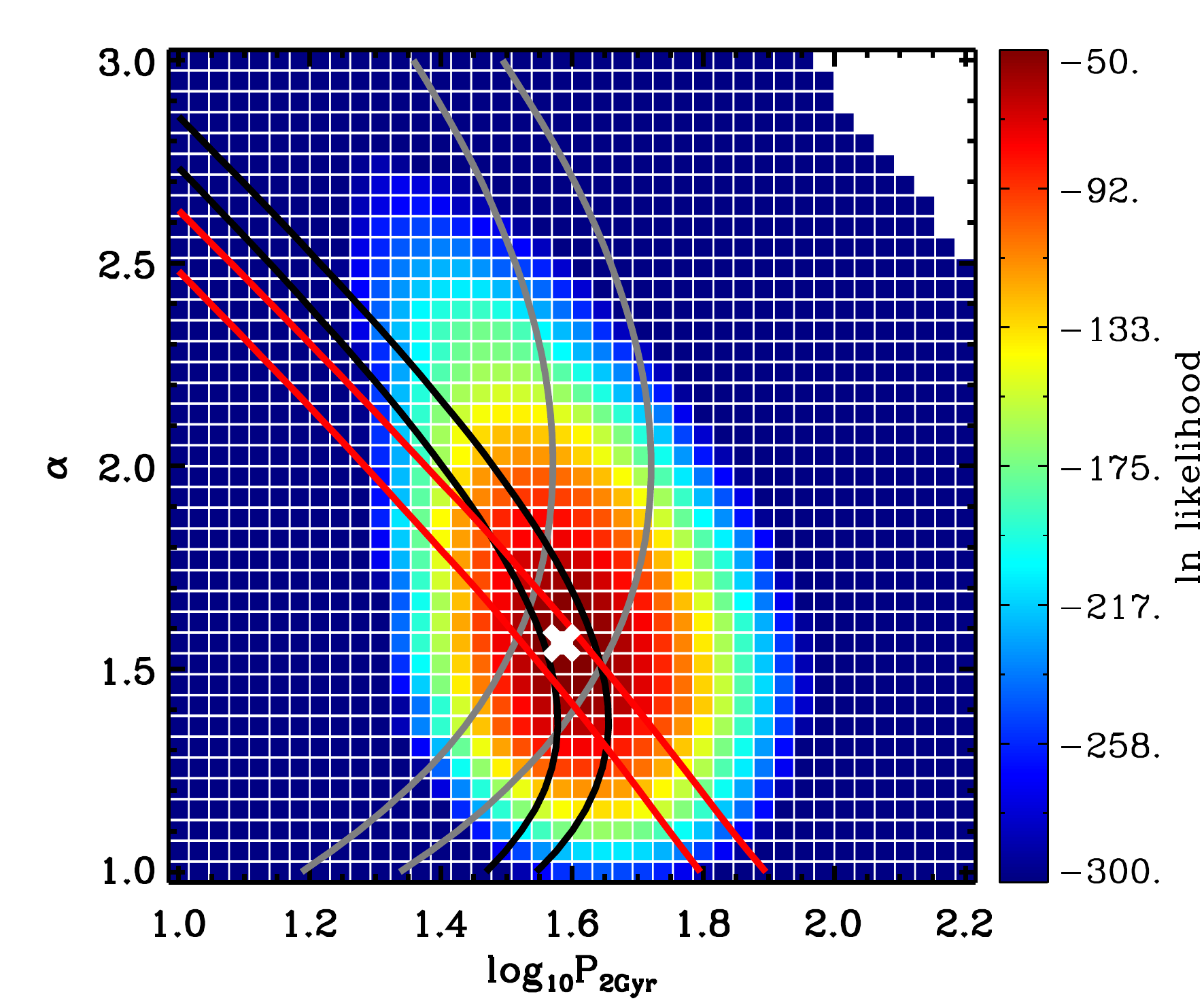}
    \end{center}
  \caption{First three panels: The $\sigma_{\rm 7}$, $\sigma_{\rm 9}$, $\sigma_{\rm 79}$ that are derived for a population of mock galaxies as a function of the assumed $P_{\rm 2Gyr}$-$\alpha$ of the PSD.   In each panel, the two tram-line contours then represent the $\pm$2$\sigma$ values of these three parameters that are observed in the (integrated) MaNGA sample of Figure \ref{fig:global}.
  In the lower right panel, the colour scale indicates 
 the likelihood function that is obtained by directly comparing the 2-dimensional probability distributions from Figure \ref{fig:example} to the observed distribution of points in Figure \ref{fig:global}. The maximum likelihood is indicated by the white cross.  The three sets of contours from each of the other three panels are copied over to this panel and reassuringly intersect at the point of maximum likelihood (see text).  
  }
  \label{fig:para_space}
\end{figure*}
        
We first consider a simple model of the PSD for $\Delta$sSFR(t), a single power-law, without 
including any intrinsic scatter ($\sigma_{\rm int}=0$).  The PSD of $\Delta$sSFR(t) 
can be written as: 
\begin{equation} \label{eq:6}
{\rm PSD}(\nu) =  P_{\nu_0} (\frac{\nu}{\nu_{0}})^{-\alpha},
\end{equation}
where the $P_{\nu_0}$ is the amplitude of PSD at $\nu_{\rm 0}$, some arbitrary reference frequency. As discussed in Section \ref{subsec:3.1}, there should be no power 
at frequencies below $\tau_{\rm Hubble}^{-1}$, and in the present work, we ignore for simplicity the power on all timescales greater than 10 Gyr, which is comparable to the lifetime of the galaxy 
population\footnote{The power between $\tau_{\rm Hubble}^{-1}$ and 0.1 Gyr$^{-1}$
is ignored in this case. While this is not a concern, we have added the 
intrinsic scatter in our main result, which would compensate this power loss. }.
According to Equation \ref{eq:5}, the value of $P_{\nu_0}$ for a general $\nu_0$
should be strongly correlated with the slope $\alpha$ of the PSD, in order that the integrals 
in Equation \ref{eq:5} converge to the observed values.  There will generally be one $\nu_0$ where this dependence vanishes, and for the current set-up, we find that  
the $P_{\nu_0}$ is not sensitive to $\alpha$ for 
$\nu_0 =$ 0.5 Gyr$^{-1}$. Therefore, we chose for convenience $\nu_0$= 0.5 Gyr$^{-1}$
throughout this work, and denote the $P_{\nu_0}$ as $P_{\rm 2Gyr}$. 
We note that the shape of the constrained PSD does not depend in any way on the choice of 
$\nu_0$ and that $\nu_0$ has no physical significance whatsoever.
%while the $P_{\nu_0}$ undoubtedly depends on the choice of $\nu_0$. 

\subsubsection{The distribution of points on the $\Delta$sSFR7-$\Delta$sSFR9 diagram}

In order to gain a first impression of how $\alpha$ and $P_{\rm 2Gyr}$ 
determine the distribution of galaxies in the $\Delta$sSFR7-$\Delta$sSFR9 
diagram, we generate mock galaxies from SFH histories constructed from PSD with different $\alpha$ and $P_{\rm 2Gyr}$. This is 
shown in Figure \ref{fig:example}.  The distribution of mock galaxies on 
the $\Delta$sSFR7-$\Delta$sSFR9 diagram can be well characterized in each case by an inclined 
two-dimensional Gaussian surface. Therefore, we directly show the 2d-Gaussian
fits of the these distributions in Figure \ref{fig:example}. 
This 2d-Gaussian distribution can also be regarded to be the probability distribution 
of galaxies on the $\Delta$sSFR7-$\Delta$sSFR9 diagram for a given $\alpha$ and $P_{\rm 2Gyr}$.

We generate the mock galaxies in 
the following way. First, we construct the time-series of 
$\Delta$sSFR(t) in logarithmic space for 10,000 mock galaxies from a PSD with 
$\alpha$ and $P_{\rm 2Gyr}$ by using a public IDL code\footnote{https://github.com/svdataman/IDL/tree/master/src}. 
The length of the time-series is 10 Gyr, which means that we ignore the power on 
timescales greater than 10 Gyr.  As in \citetalias{Wang-19b}, the time resolution 
is set to be 1 Myr, which is smaller than the SFH timescale traced by H$\alpha$ emission, 
and also smaller than the free-fall timescale of molecular clouds \citep{Murray-10,
Hollyhead-15, Freeman-17}. 
Then, based on this $\Delta$sSFR(t), 
we produce the sSFH(t) of each mock galaxy by adding the cosmic evolution of the mean sSFR of the SFMS. 
We adopt the evolution of SFMS from \cite{Lilly-16}, which can be written as: 
\begin{equation} \label{eq:7}
    {\rm sSFR}(z) = 0.07\times(1+z)^2 \ {\rm Gyr}^{-1} 
\end{equation}
This change with redshift is broadly consistent with the observational 
estimates of this quantity\footnote{In principle, the evolution of the SFMS also shows
a very week dependence of the stellar mass, which is negligible at a timescale of 800 Myr.} \citep[e.g.][]{Pannella-09, Stark-13}. 
This sSFR(t) is then converted to the SFR(t) by direct integration, from which the $\Delta$sSFR7 and $\Delta$sSFR9 may be calculated, using the {\it current} (i.e. final) stellar mass of the galaxies, 
exactly as for the observations. 
%In this process, we assume a fixed final stellar mass ($z=0$) to be $10^{10.5}$\msolar\ for all the mock galaxies, which is comparable to the median stellar mass of our galaxy sample.  We argue that our result does not depend on the assumption of the final stellar mass, since the evolution of SFMS we adopted does not depend on the stellar mass. 

As can be seen in Figure \ref{fig:example}, at a fixed $\alpha$, the overall
distribution of mock galaxies gets broader with increasing $P_{\rm 2Gyr}$ (left to right along each row), while 
the overall shape of the distribution remains unchanged, i.e. the correlation between
$\Delta$sSFR7 and $\Delta$sSFR9 is not changed. 
This can be understood,
since the $P_{\rm 2Gyr}$ determines the overall power, and varying only $P_{\rm 2Gyr}$
would result in the same (logarithmic) change of $\Delta$sSFR7 and $\Delta$sSFR9. 
At a fixed $P_{\rm 2Gyr}$, the slope of the distribution more nearly approaches  diagonal 
with increasing $\alpha$, and the distribution appears to stretch along the major-axis. The $\Delta$sSFR7 and $\Delta$sSFR9 become 
more and more correlated with increasing $\alpha$. 
Since $\alpha$ determines the power distribution at low and high frequency, 
the larger $\alpha$ reduces the variation on short timescales, which 
causes a stronger correlation between $\Delta$sSFR7 and $\Delta$sSFR9. 

Figure \ref{fig:example} shows how the $\alpha$ and $P_{\rm 2Gyr}$
can in principle be determined by matching the distribution of observed $\Delta$sSFR7 and $\Delta$sSFR9 of the galaxies with these computed probability distributions (see Figure \ref{fig:global} and Figure \ref{fig:data_resolved}). 

\subsubsection{Constraints on the PSD using $\sigma_7$, $\sigma_9$ and $\sigma_{79}$}

The distribution of observed $\Delta$sSFR7, $\Delta$sSFR9 points can of course be reduced to simpler quantities, in particular the 
observed dispersions $\sigma_7$, $\sigma_9$ and $\sigma_{79}$.  The dependence of these on $\alpha$ and $P_{\rm 2Gyr}$ can be seen as follows. 
Figure \ref{fig:para_space} shows with color-coding the $\sigma_7$, $\sigma_9$ and 
$\sigma_{79}$ values across $\alpha$-$P_{\rm 2Gyr}$ parameter space. At each given $\alpha$ and $P_{\rm 2Gyr}$,
we generate 10,000 mock galaxies in the same way as above, and then calculate the
$\sigma_7$, $\sigma_9$ and $\sigma_{79}$ for this model population. 

The black, red and gray contours in the first three panels represent the observed $\sigma_7=0.338\pm0.015$ dex, 
$\sigma_9=0.265\pm0.014$ dex, and $\sigma_{79}=0.233\pm0.020$ dex, obtained from the observational data in Figure \ref{fig:global}, with the uncertainties calculated using the bootstrap 
method to be $\pm 2 \sigma$ intervals. For this initial, illustrative, exercise, we do not consider the contribution of the observational uncertainties to the observed $\sigma_7$, $\sigma_9$ and $\sigma_{79}$.
We also stress that we are here using the ``integrated'' sample of Figure \ref{fig:global} only for illustrative purposes, ignoring the possible 
non-ergodicity for galaxies of different masses that we discussed above. 

As shown in Figure \ref{fig:para_space}, the $\sigma_7$ and $\sigma_9$ are sensitive to 
both $\alpha$ and $P_{\rm 2Gyr}$, while the $\sigma_{79}$ appears to be more sensitive to $P_{\rm 2Gyr}$
than $\alpha$.  At fixed $\alpha$, both $\sigma_7$, $\sigma_9$ and $\sigma_{79}$
are increasing with $P_{\rm 2Gyr}$, while at fixed $P_{\rm 2Gyr}$, they are not monotonic
functions of $\alpha$. The $\sigma_9$ increases for $\alpha>1$. At $\alpha<1$, the contours of $\sigma_9$
show similar curved features as those of $\sigma_7$ or $\sigma_{79}$. 
Although the curved features correspond to different $\alpha$ for $\sigma_7$, $\sigma_9$ 
and $\sigma_{79}$, they are all due to the fact that 
the dominant variations change from the short timescale variations 
to long timescale variations when $\alpha$ is increased from 0 to 3. 

The three sets of contours in the first three panels are transcribed onto the bottom right panel of Figure \ref{fig:para_space}.  Reassuringly, they broadly intersect at one point. However, it should be remembered
that the $\sigma_7$, $\sigma_9$ and $\sigma_{79}$ are not fully independent and in principle,
any two of them can give constrains of the $\alpha$ and $P_{\rm 2Gyr}$. As an aside, however, this concordance  also suggests
that our measurements of SFR79 are sensible, since problematic measurements of SFR79 would 
not necessarily allow the three contours intersect at one point.  This 
intersection point corresponds to the location of the preferred $\alpha$ and $P_{\rm 2Gyr}$, at least for the illustrative integrated sample of Figure \ref{fig:global} and neglecting for the contribution of any intrinsic scatter.

\subsubsection{The constraint on the PSD using the likelihood method}

Having obtained the probability distribution of mock galaxies on the $\Delta$sSFR7-$\Delta$sSFR9 
diagram for a given $\alpha$ and $P_{\rm 2Gyr}$ (see Figure \ref{fig:example}), we can calculate the value of the likelihood based on the observed galaxy distribution in Figure \ref{fig:global}.  For consistency with the previous sub-section, we again ignore here the effect of observational errors on the observed galaxy distribution.
The bottom right panel of \ref{fig:para_space} shows the color-coding of the likelihood values 
on the $\alpha$-$P_{\rm 2Gyr}$ parameter space.

The point of 
the maximum likelihood, marked by the white cross, matches very well the intersection of the plotted contours for the constraints from $\sigma_7$, $\sigma_9$, and $\sigma_{79}$ individually.
We conclude that using the likelihood method and using the scatter of SFR7, SFR9 and SFR79 
to constrain the PSD give very consistent results. 
Figure \ref{fig:example} and Figure \ref{fig:para_space} illustrate in detail how the PSD can be constrained using the distribution of galaxies on the 
$\Delta$sSFR7-$\Delta$sSFR9 diagram. 

However, the above two methods to constrain the PSD are clearly not identical. By only using the 
values of $\sigma_7$, $\sigma_9$ and $\sigma_{79}$, we would lose the information of the detailed
distribution of galaxies.  The likelihood method has made use of the full distribution of galaxies 
on the $\Delta$sSFR7-$\Delta$sSFR9 diagram. As noted, in both methods, we did not so far consider the uncertainties of 
$\Delta$sSFR7 and $\Delta$sSFR9, but these could be easily included as a convolution of the predicted probability distribution by an appropriate kernel. 
Therefore, in the main result of this work (Section \ref{sec:4} and \ref{sec:5}), 
we will use the Markov chain Monte Carlo (MCMC) method
to constrain the PSD of $\Delta$sSFR history, including the uncertainties of 
$\Delta$sSFR7 and $\Delta$sSFR9.   

\subsection{Considering the intrinsic scatter} \label{subsec:3.4}

\begin{figure*}
  \begin{center}
    \includegraphics[width=0.45\textwidth]{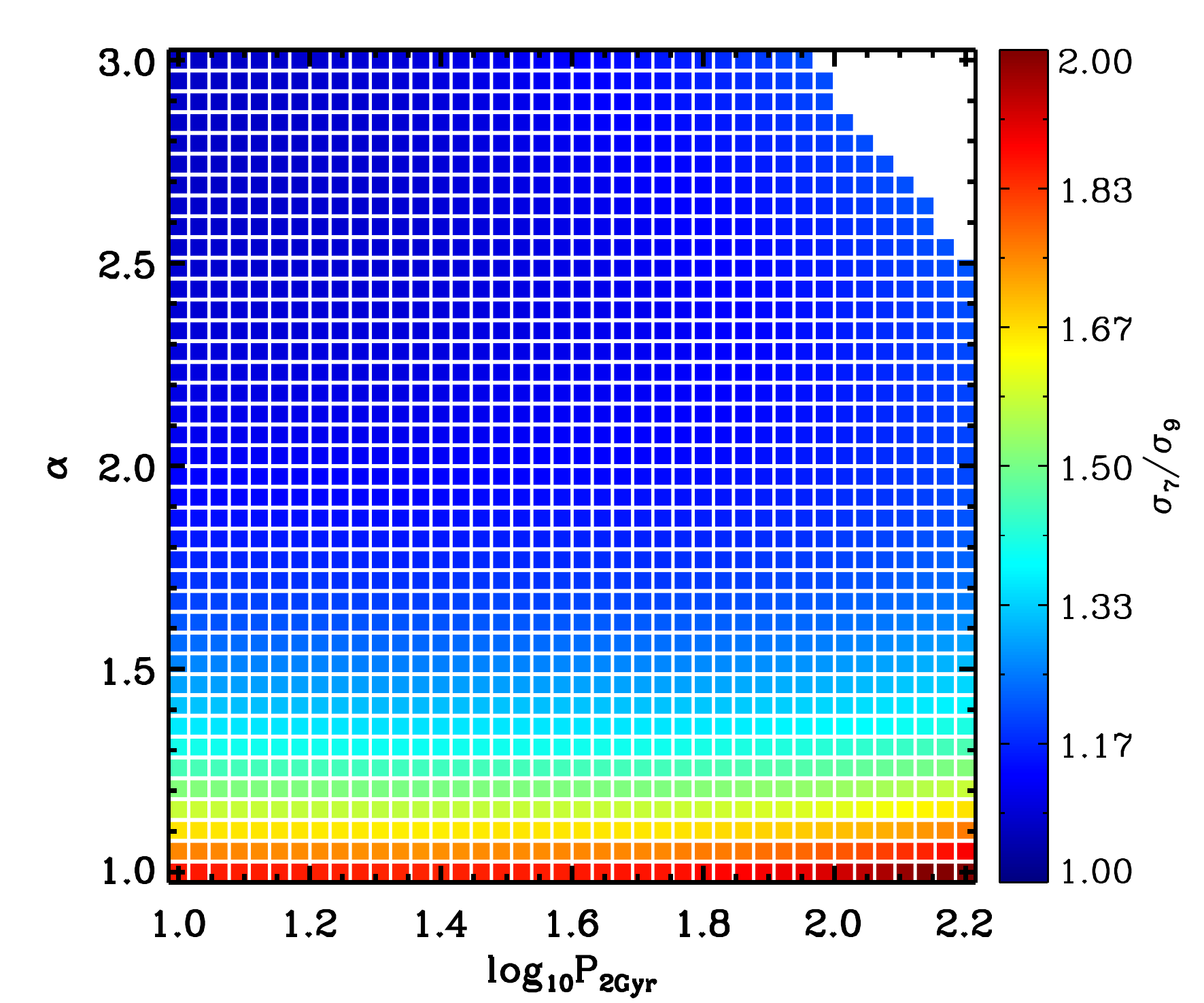}
    \includegraphics[width=0.45\textwidth]{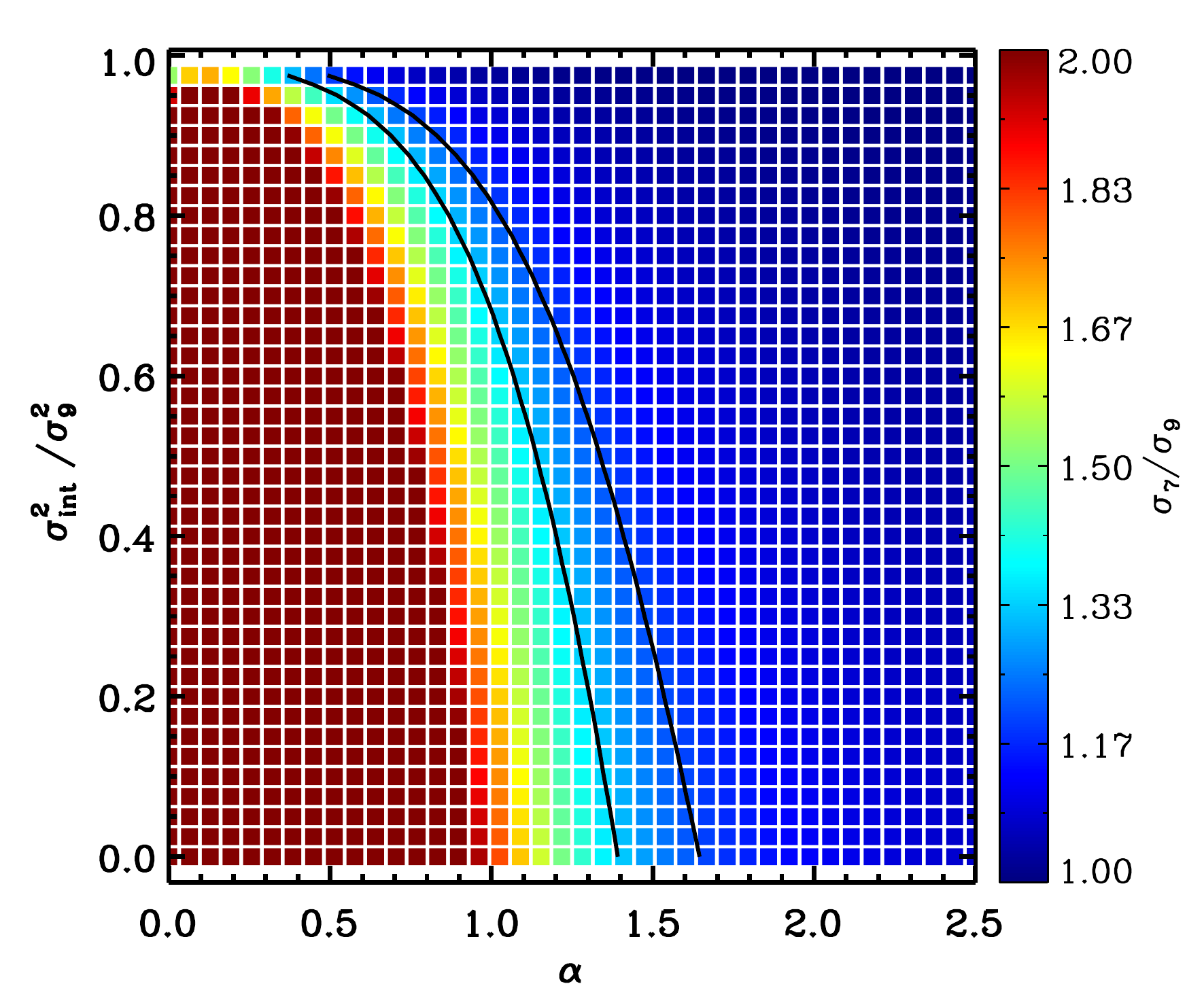}
    \end{center}
  \caption{Left panel: The ratio $\sigma_7/\sigma_9$ across the 
  $P_{\rm 2Gyr}$-$\alpha$ parameter space. Right panel: The ratio $\sigma_7/\sigma_9$
  in the $\sigma_{\rm int}^2/\sigma_9^2$-$\alpha$ diagram.  The black lines 
  correspond to the 2$\sigma$ value of $\sigma_7/\sigma_9 = 1.30\pm0.07$ that is measured from the (integrated) MaNGA data from Figure \ref{fig:global}, calculated using the
  bootstrap method. 
  }
  \label{fig:para_int}
\end{figure*}

In Section \ref{subsec:3.3}, we did not include at all the possible intrinsic scatter $\sigma_{\rm int}$ of the SFMS that we introduced earlier. 
As discussed in Section \ref{subsec:3.2}, the existence of some
intrinsic scatter is quite plausible. Idealising the intrinsic scatter as a time-invariant offset in the sSFR,
the effect of it 
will be to broaden the distribution of $\Delta$sSFR7 and $\Delta$sSFR9 by the same amount. Therefore,
the intrinsic scatter would stretch the distribution of mock galaxies along the diagonal direction (see Figure \ref{fig:example}).
The $\sigma_{79}$ does not change with the addition of such intrinsic scatter, emphasizing
the importance of our change parameter, SFR79. 
This further means that inclusion of the intrinsic scatter will shift the required PSD parameters along the locus of constant 
$\sigma_{79}$ in the $\alpha$-$P_{\rm 2Gyr}$ plane in the bottom left panel of Figure \ref{fig:para_space}. We will discuss this 
further in Section \ref{sec:4}. 

The left panel of Figure \ref{fig:para_int} shows with color-coding of the ratio $\sigma_7$/$\sigma_9$
across the $\alpha$-$P_{\rm 2Gyr}$ space, again constructed assuming $\sigma_{\rm int}=0$. 
As can be seen, for $\alpha>2$, the $\sigma_9$ is quite close to $\sigma_7$, because the 
short timescale variations ($<$800 Myr) are completely negligible with respect to longer timescale variations. 
It can be seen that the ratio of $\sigma_7$ to $\sigma_9$ only depends on the slope of the PSD, 
rather than the overall amplitude. This is expected, since $\alpha$
determines the distribution of the power on different timescales, as discussed in Section 
\ref{subsec:3.3}.  In other words, the slope of the PSD is determined by $\sigma_7$/$\sigma_9$ (with 
$\sigma_{int}=0$).  When we add in the possible contribution of intrinsic scatter, the relevant ratio determining $\alpha$
becomes $\sqrt{\sigma_7^2-\sigma_{\rm int}^2}/\sqrt{\sigma_9^2-\sigma_{\rm int}^2}$\footnote{Actually, 
it is not exactly $\sqrt{\sigma_7^2-\sigma_{\rm int}^2}/\sqrt{\sigma_9^2-\sigma_{\rm int}^2}$, 
because the $\sigma_{\rm int}$ has a slightly different impact on $\sigma_7$ and $\sigma_9$, 
due to the definition of $\sigma_7$ and $\sigma_9$. However, this difference is negligible
given the current sSFR$\ll$1 Gyr$^{-1}$. }. 

The value of $\sigma_{\rm int}$ is not completely unconstrained since it must be less than the observed $\sigma_9$. 
Therefore, we show in the right panel of Figure \ref{fig:para_int} the color-coding of the final observed $\sigma_7/\sigma_9$ on the $\sigma_{\rm int}^2/\sigma_9^2$-$\alpha$ parameter space, at an arbitrary fixed $P_{\rm 2Gyr}$ (note that this plot does not depend
on the value of $P_{\rm 2Gyr}$, because $\sigma_7/\sigma_9$ does not depend on $P_{\rm 2Gyr}$). 
Here we use $\sigma_{\rm int}^2/\sigma_9^2$, rather than $\sigma_{\rm int}/\sigma_9$, because
$\sigma_{\rm int}^2/\sigma_9^2$ directly characterizes the fraction of the variance contributed by 
intrinsic scatter for the variations of $\Delta$sSFR9. 
The black lines indicate the locus of observed $\sigma_7/\sigma_9 = 1.30\pm0.07$ that is obtained from
the observation. 

The right panel of Figure \ref{fig:para_int} directly shows how changes in the assumed contribution of $\sigma_{\rm int}$ change the power-law index $\alpha$ of the returned PSD. As shown, the required $\alpha$ decreases slowly with 
increasing $\sigma_{\rm int}$ at first, but then decreases rapidly when $\sigma_{\rm int}$ 
approaches $\sigma_9$. Actually, with increasing $\sigma_{\rm int}$, the power 
between 1/800 Myr$^{-1}$ and 1/10 Gyr$^{-1}$ would decrease, and the power 
below the frequency of 1/800 Myr$^{-1}$ would increase.  This is why 
the required slope of the PSD gets flatter with increasing $\sigma_{\rm int}$. 
In the extreme case, when $\sigma_{\rm int}$ is close to $\sigma_9$, the long timescale 
variations are very small, and almost all the variations must be contributed by  
short timescale modes. This is the reason why the $\alpha$ decreases rapidly when
$\sigma_{\rm int}$ approaches $\sigma_9$. 

It is clear from Figure \ref{fig:para_int} that ignoring the possible contribution of $\sigma_{\rm int}$ 
will lead to an over-estimate of the slope of the PSD \citep[c.f.][]{Caplar-19}. 

\section{Application to the integrated sample}   \label{sec:4}

In order to explore how the PSD can be determined from the data, we focus in this Section on the integrated sample shown in Figure \ref{fig:global}, in which a single point is plotted for each galaxy (based on the spectra integrated out to \re), and galaxies of all stellar masses are considered together.  As remarked above, it is by no means clear that galaxies of different masses can be considered an ergodic sample, and this section should therefore be regarded as being of heuristic interest.  The final analysis of the spatially resolved data, in which galaxies of different mass will be considered separately, will be presented in the next section. 

\subsection{MCMC analysis} \label{subsec:4.1}

As we discussed in Section \ref{subsec:3.3}, we will use the MCMC method to constrain
the parameters of the PSD.  The essential role of MCMC is to define the likelihood function. 
% SJL  XXXXXX  Did not fully understand.
In this work, we obtain the value of the likelihood function
following the same approach described earlier in Section \ref{subsec:3.3}. 
In brief, for a given PSD, we generate 10,000 mock galaxies, and further 
obtain the probability distribution of galaxies on the $\Delta$sSFR7-$\Delta$sSFR9
diagram. Unlike previously in Section \ref{subsec:3.3}, we now explicitly include the 
observational uncertainties in the location of individual galaxies in the $\Delta$sSFR7-$\Delta$sSFR9
diagram by convolving the probability distribution of the mock galaxies with the typical 
uncertainties of galaxies in $\Delta$sSFR7 and $\Delta$sSFR9
(see the blue ellipse in Figure \ref{fig:global} and Figure \ref{fig:data_resolved}). 
Combining with the observed distribution of galaxies 
(Figure \ref{fig:global} and Figure \ref{fig:data_resolved}),  we therefore obtain 
the likelihood value for a given PSD. 

%  --  Detail setting in MCMC method ...
%  --  The basic performance of MCMC running ...

We take advantage of a public {\tt python} code: {\tt emcee}\footnote{https://github.com/dfm/emcee}, 
which has several advantages over traditional MCMC sampling methods and 
excellent performance as measured by the auto-correlation time \citep{Foreman-Mackey-13}. 
In practice, we assume a uniform prior of $\alpha$ in the range of [$-$2.0,5.0], and also 
a uniform prior of $P_{\rm 2Gyr}$ in the range of [0,10000]. 
The setting of the wide range of $\alpha$ and $P_{\rm 2Gyr}$ enables the MCMC to 
search for the best-fitting parameters. 
The initial values of $P_{\rm 2Gyr}$ and $\alpha$ are given according to Figure \ref{fig:para_space},
with small perturbations for each walker. 
To explore the parameter space, we set up 6 walkers and 4000 steps for each walker. 
Although it seems that the number of walkers and steps are not very large, we find 
that the burn-in phase is usually less than 1000 steps and the chains 
quickly converge. This is because the likelihood function 
is a single peaked function in the parameter space (see Figure \ref{fig:para_space}).  
Throughout this work, we keep the same settings for all the MCMC runs. 

%  --  Why we do not put the $\sigma_{int}$ as another free parameter 
%    in the MCMC fitting ... 
%      -- longer chains, and time consuming ... 
%      -- over-fitting problems ... 
%      -- no possible for the resolved results ... because the intrinsic scatter should be 
%         the same for different radial bins, at given stellar mass ... 

In Section \ref{subsec:3.4}, we discussed the impact of any intrinsic scatter 
in determining the slope of PSD. In principle, we could add the intrinsic scatter 
as another free parameter in the MCMC analysis.  However, as discussed in Section \ref{subsec:3.4},
the $\sigma_7$, $\sigma_9$ and $\sigma_{79}$ are not fully independent, therefore we could 
not in principle give a good constrain for the model of PSD with more than two free 
parameters.  This means that $\alpha$ and $\sigma_{\rm int}$ are highly degenerate (also as seen on
the right panel of Figure \ref{fig:para_int}).  
Furthermore, when we come to determining the PSD for the 
resolved datasets (i.e. at different radii and at different galactic masses), it is not clear that we should allow $\sigma_{\rm int}$ to vary with radius within a particular mass bin. 
%because the resulting $\sigma_{\rm int}$ are not likely to be the same for galaxies at different galactic radii, but should be the same under our frame of the galaxy evolution that the star formation in galaxies are primarily governed by the global cold gas inflow.  
In order to avoid the possible over-fitting problem and to avoid the possible  
inconsistency of $\sigma_{\rm int}$ at different galactic radii, we will therefore constrain 
the PSD of $\Delta$sSFR(t) for each of a discrete set of $\sigma_{\rm int}$.  

\subsection{The returned PSD without $\sigma_{\rm int}$} \label{subsec:4.2}

\begin{figure*}
  \begin{center}
    \includegraphics[width=0.56\textwidth]{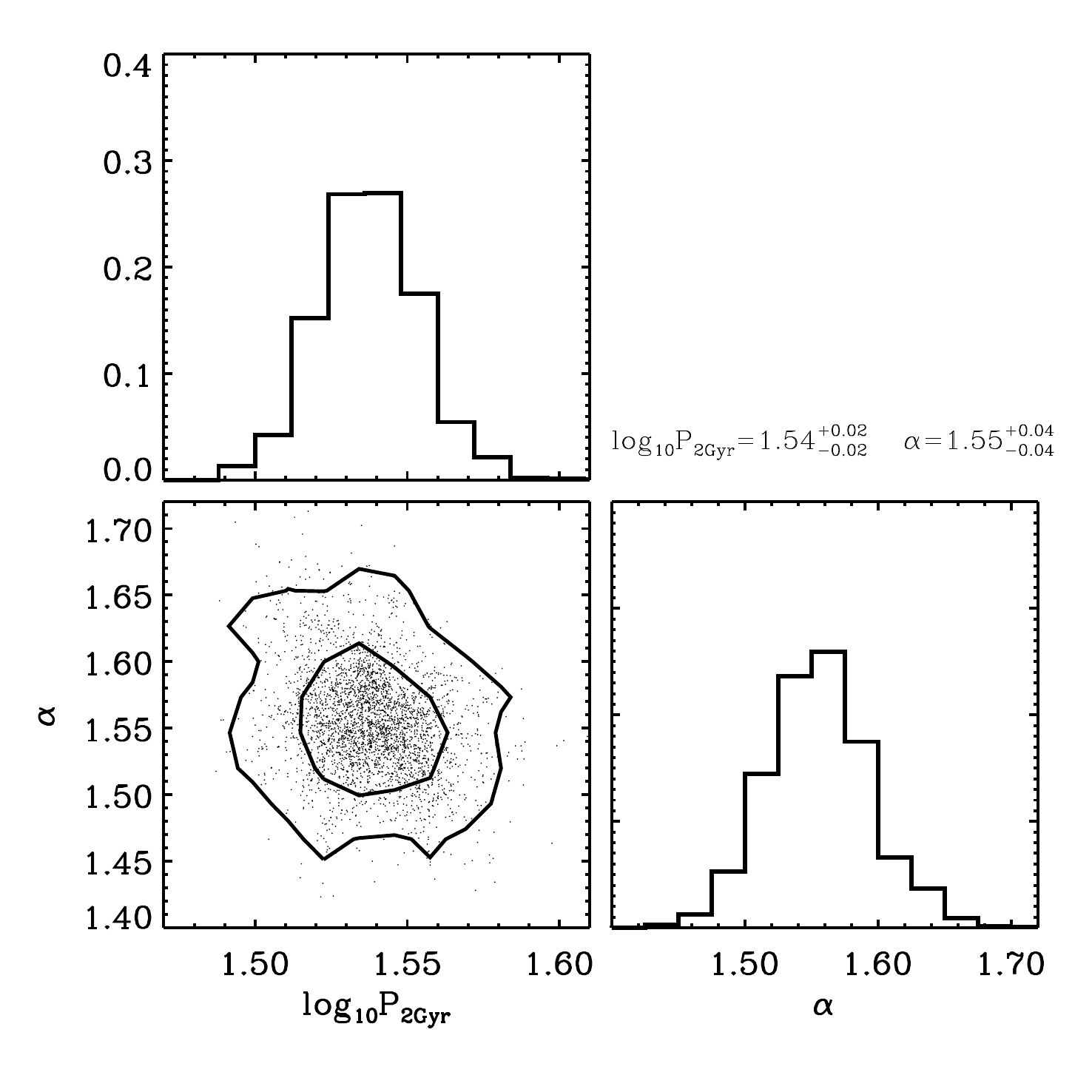}
    \includegraphics[width=0.37\textwidth]{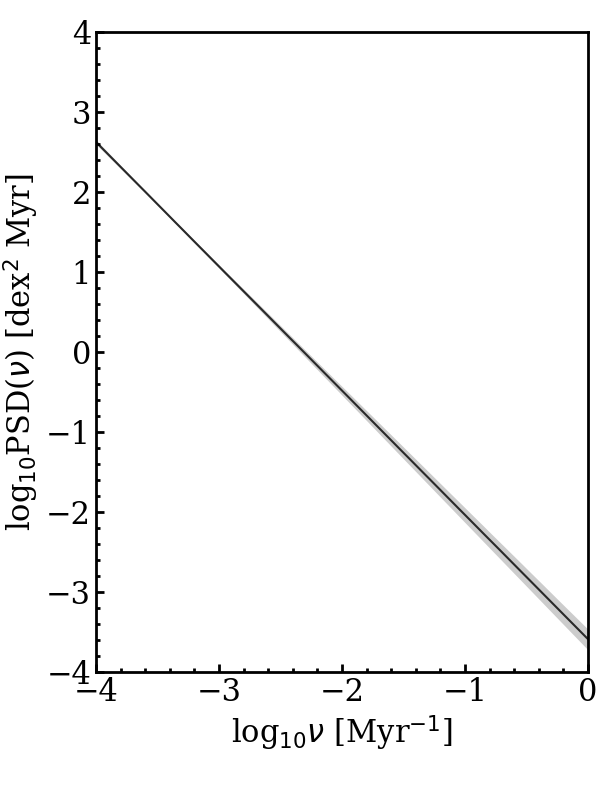}
   \end{center}
  \caption{Left group of panels: the distribution of $\alpha$ and $P_{\rm 2Gyr}$ returned 
  by the MCMC analysis. In the bottom left panel, the distribution of the small dots shows the 
  joint probability distribution of $\alpha$ and $P_{\rm 2Gyr}$. The best-fit parameters 
  as well as their uncertainties are denoted on the plot. 
  Right panel: the returned PSD and its 1$\sigma$ uncertainty.}
  \label{fig:psd_2p_global}
\end{figure*}

Based on the approach above, we can obtain the PSD as well as the 
uncertainties for any given $\sigma_{\rm int}$ ($<\sigma_9$). 
We first look at the simplest case, with  
$\sigma_{\rm int}=0$. Again for illustration purposes, we use the full mass sample to 
constrain the PSD of $\Delta$sSFR(t). We note that the PSD may well 
vary with stellar mass, which will be investigated in the next section. 
The left panel of Figure \ref{fig:psd_2p_global} shows 
the constrained probability distribution of the parameters based on our MCMC analysis,
and the right panel of Figure \ref{fig:psd_2p_global} shows the returned PSD
of $\Delta$sSFR(t) and its 1$\sigma$ uncertainty. 

As shown, both $\alpha$ and $P_{\rm 2Gyr}$ are very well constrained with $\sigma_{\rm int} = 0$. 
In the joint probability distribution of $\alpha$ and $P_{\rm 2Gyr}$, we do 
not see a strong correlation between the two parameters, as expected given the reason for choosing 2 Gyr to define $\nu_0$ in Equation \ref{eq:6}. Choosing a different $\nu_0$ would certainly change the constrained probability 
distribution of $P_{\nu_0}$ and $\alpha$, but would not change the returned PSD plotted in the right 
panel of Figure \ref{fig:psd_2p_global}.  

In \citetalias{Wang-19b}, we applied an arbitrary correction to eliminate the small dependence 
of SFR79 on $\Sigma_*$ (see section 3.3 of \citetalias{Wang-19b} for details). To check 
if this correction has any impact on the present analysis, we repeat it without this correction applied, and find very little effect on the returned PSD. 
In the remainder of this work, 
we only present the results obtained with this correction, which are thereby completely consistent
with the data we used in \citetalias{Wang-19b}.  

The returned parameters from the MCMC analysis are labeled in Figure \ref{fig:psd_2p_global}: 
$\log_{10}P_{\rm 2Gyr} = 1.54\pm 0.02$, and $\alpha=1.55\pm 0.04$, which is quite 
similar to the ones measured in Section \ref{subsec:3.3}. This indicates that the 
MCMC works well in exploring the parameter space. It also suggests that the 
inclusion of the observational uncertainty on the distribution of data points only has a very minor effect 
in determining the PSD.  

The returned slope of the PSD is $\sim$1.5 for $\sigma_{\rm int}=0$.  This 
indicates that the variance of $\Delta$sSFR(t) is dominated by the longer
timescale variations ($>$1 Gyr), since $\alpha < 1$ would be required to produce a $\nu$PSD (the contribution to the variance per log interval of frequency) that increased to high frequencies.

\subsection{The returned PSD with $\sigma_{\rm int}$} \label{subsec:4.3}

% ----  sigma_int ----- 

\begin{figure*}
  \begin{center}
    \includegraphics[width=0.38\textwidth]{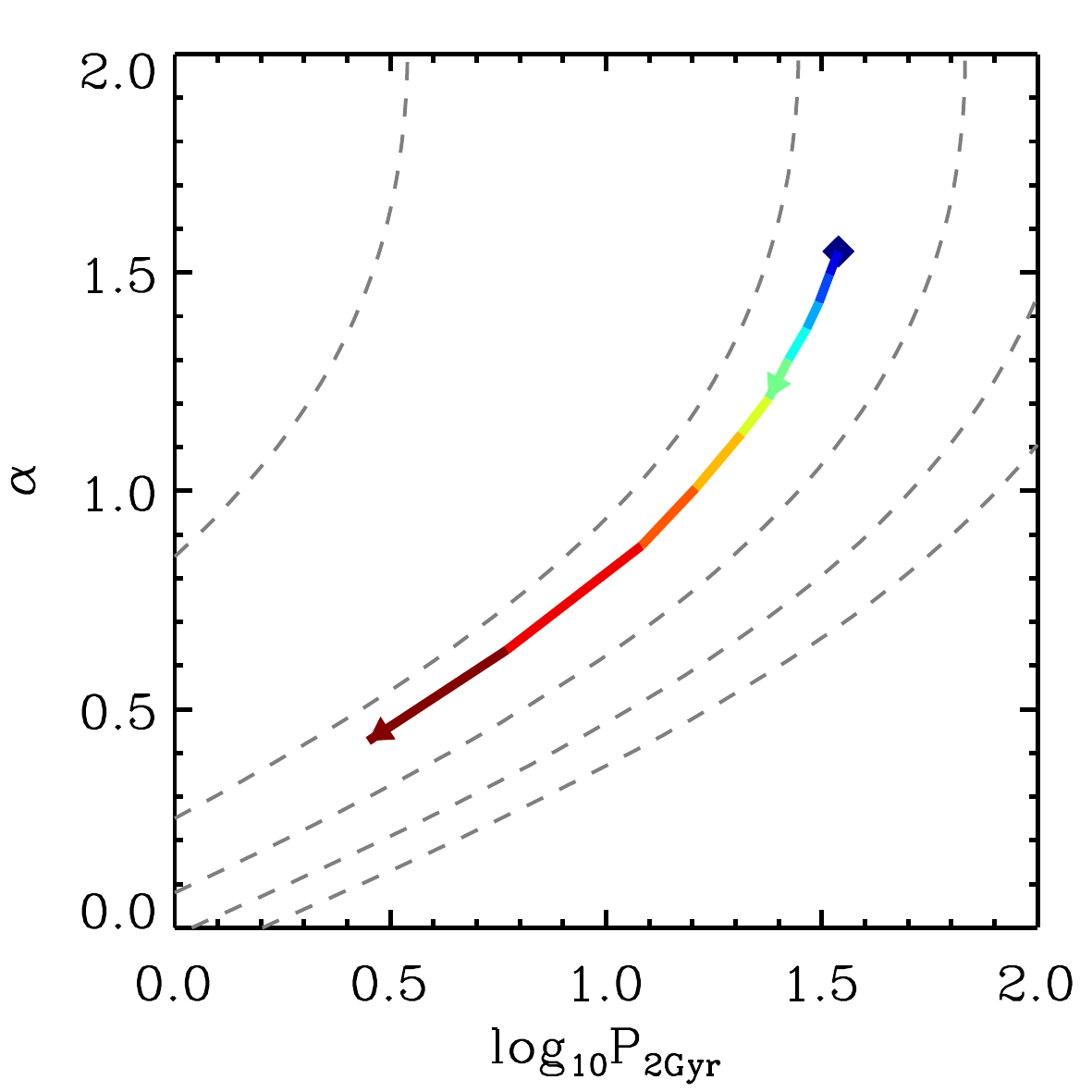}
    \includegraphics[width=0.57\textwidth]{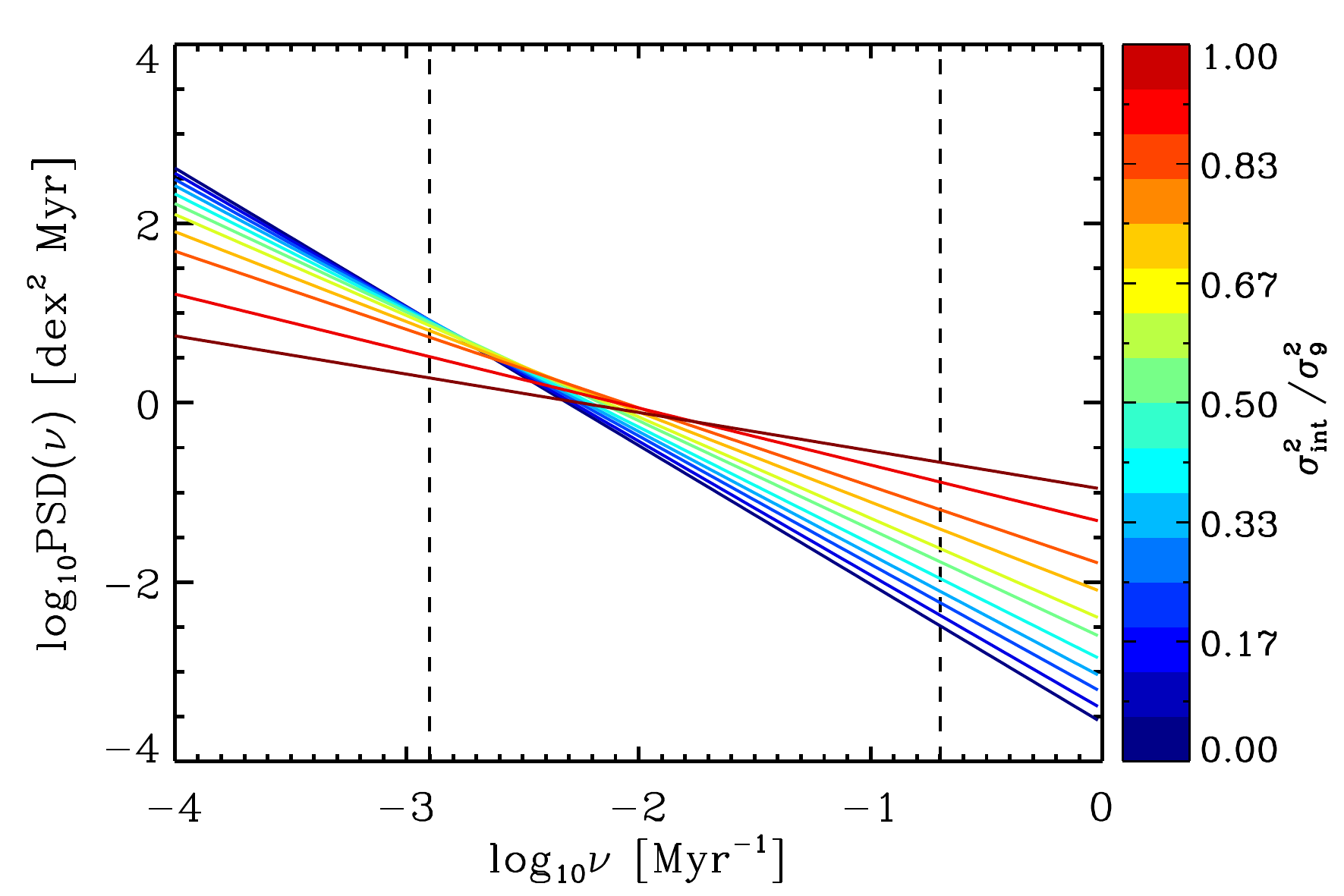}
    \end{center}
  \caption{Left panel: the track of the returned parameters on the $\alpha$-$P_{\rm 2Gyr}$ diagram 
  when different values of the intrinsic scatter are assumed.  In the left panel, 
  the blue diamond indicates the best constrained parameters with $\sigma_{\rm int}=0$, 
  and the track to the lower left shows the effect of increasing $\sigma_{\rm int}$.  
  The dashed lines are contours of constant $\sigma_{\rm 79}$ in 
  the parameter space, taken from Figure \ref{fig:para_space}, and correspond to 0.1, 0.2, 0.3 ,0.4, and 0.5 dex 
  from top-left to bottom-right, respectively (but now with the addition of the observational dispersion of 0.076 dex in SFR79).
  As expected, the effect of increasing $\sigma_{\rm int}$ moves the returned parameters along a track of constant $\sigma_{\rm 79}$.  Along the track, the assumed $\sigma_{\rm int}^2$ as a fraction of the observed $\sigma_9^2$ is indicated by the color scale to the right.
  Right panel: The returned PSDs for different assumed $\sigma_{\rm int}$, as also indicated 
  by the same color scale to the right.  The two dashed vertical lines indicate  
  the frequencies of 1/5 Myr$^{-1}$ and 1/800 Myr$^{-1}$. 
  }
  \label{fig:psd_2p_3p}
\end{figure*}

Now we look at the cases with $\sigma_{\rm int}>0$. As mentioned in Section \ref{subsec:3.4},
the $\sigma_{\rm int}$ must be less than or equal to $\sigma_9$. 
We therefore explore the PSD of $\Delta$sSFR(t) assuming a set of $\sigma_{\rm int}$: $\sigma_{\rm int}^2/\sigma_9^2$= 0.1, 0.2, 0.3, 0.4, 0.5, 0.6, 0.7, 0.8, 0.9, and 1.0.  
%As mentioned in Section \ref{subsec:3.4}, $\sigma_{\rm int}^2/\sigma_9^2$, rather than $\sigma_{\rm int}/\sigma_9$,  characterizes the fraction of variations contributed by intrinsic scatter for the variation of $\Delta$sFR9. 
At each $\sigma_{\rm int}$, 
%in the same approach of Section \ref{subsec:4.1}, 
we run the MCMC to determine the joint probability distribution of $\alpha$ and $P_{\rm 2Gyr}$. 

The left panel of Figure \ref{fig:psd_2p_3p} shows the change of returned parameters 
with increasing $\sigma_{\rm int}$.  The dashed lines represent the lines of constant $\sigma_{79}$
in the parameter space (also see Figure \ref{fig:para_space}). The right panel of 
Figure \ref{fig:psd_2p_3p} shows the change in the returned PSD with increasing 
$\sigma_{\rm int}$.  In both panels, the different colors represent different $\sigma_{\rm int}^2/\sigma_9^2$. 

As shown, with increasing $\sigma_{\rm int}$, both $\alpha$ and $P_{\rm 2Gyr}$ become 
smaller, following the track of constant $\sigma_{79}$. 
%As discussed in Section \ref{subsec:3.3}, the intrinsic scatter does not contribute to the dispersion of SFR79. 
Consistent with the analysis in \ref{subsec:3.3}, the $\alpha$ and $P_{\rm 2Gyr}$ 
decrease slowly with increasing $\sigma_{\rm int}$ at $\sigma_{\rm int}^2/\sigma_9^2<$0.5, 
and then decrease much more rapidly at $\sigma_{\rm int}^2/\sigma_9^2>$0.5.  
More specifically, the slope of the PSD decreases from $\sim$1.5 to 0.3 when
$\sigma_{\rm int}$ increases from 0 to $\sigma_9$.
At $\sigma_{\rm int}^2/\sigma_9^2 \sim$ 0.8, the slope of the PSD reaches 1 and the
%The result in Figure \ref{fig:psd_2p_3p} indicates that the dominated variation of $\Delta$sSFR(t) changes
dominant timescale shifts from long timescales to short timescales. 

A similar result can also be seen in the right-hand panel of Figure \ref{fig:psd_2p_3p}. 
The PSD gets flatter with increasing $\sigma_{\rm int}$
%The variation of PSD at $\sigma_{\rm int}^2/\sigma_9^2<$0.5 is much less significant
especially at $\sigma_{\rm int}^2/\sigma_9^2>$0.5. 
It is noticeable that, for $\sigma_{\rm int}^2/\sigma_9^2<$0.5, all the returned PSDs 
%of different $\sigma_{\rm int}$ 
intersect at a single point, at $\sim$400 Myr. The effect of variations in $\sigma_{\rm int}$ is to change the slope of the returned PSD about this point. The amplitude of the PSD at this particular timescale can therefore be determined more or less independent of $\sigma_{\rm int}$ for all $\sigma_{\rm int}^2/\sigma_9^2<$0.5. This will be used below.

%Although the PSDs obtained by MCMC are in good agreement with the analysis in Section \ref{subsec:3.3} and \ref{subsec:3.4}, 

We cannot know
the value of $\sigma_{\rm int}$ from the data itself, nor do we know of a reliable way to determine it observationally by any other means. Therefore,
we are unable to say definitively from the present data whether the variation of $\Delta$sSFR(t)
is dominated by short timescale variations or long timescale ones.
%However, we argue that very large $\sigma_{\rm int}$ is not very likely. 
If the $\sigma_{\rm int}$ is large, i.e. close to the observed $\sigma_9$, then the variation of $\Delta$sSFR(t) 
is contributed almost entirely by the shortest timescale variations, with almost no
power at timescale between 1 Gyr and 10 Gyr.  This extreme situation seems to us unlikely 
to be real. In addition, it is inconsistent with 
the result from the EAGLE \citep[Evolution and Assembly of GaLaxies and their Environments;][]{Schaye-15} simulations, where the variation of SFR is 
significant at timescale between 1 Gyr and 10 Gyr \citep{Matthee-19}. 

Therefore, we suspect that the $\sigma_{\rm int}$ will not be close to $\sigma_9$ and, in the present work, we will therefore focus on the results that are obtained with $\sigma_{\rm int}^2/\sigma_9^2<$0.5,
i.e. $\sigma_{\rm int}<0.7\sigma_9$. This is equivalent to demanding that at least half of the observed variation in the SFMS (measured on Gyr timescales) comes from genuine temporal variations on timescales below 10 Gyr.  With this assumption, we can then conclude that the slope of the (single power-law) PSD is constrained to be between 1.2-1.5 and that the PSD at 400 Myr (the apparent pivot point in the right
panel of Figure \ref{fig:psd_2p_3p}) can be well determined .  

\section{Results for different masses and radii} \label{sec:5}

In the previous section, we constrained the PSD of $\Delta$sSFR(t) by using an MCMC analysis of the distribution of points from the integrated sample on the $\Delta$sSFR7-$\Delta$sSFR9 diagram (from Figure \ref{fig:global}).  We now turn to consider the constraints on the PSDs from each of the 25 panels of Figure \ref{fig:data_resolved}, in which $\Delta$sSFR7 and $\Delta$sSFR9 are determined separately in five bins of relative radius for each of five bins of galactic stellar mass (see Figure 
\ref{fig:data_resolved}), and then compare the resulting PSDs.

\subsection{The PSD of $\Delta$sSFR(t) for spatially-resolved dataset} \label{subsec:5.1}

\begin{figure*}
  \begin{center}
    \includegraphics[width=0.99\textwidth]{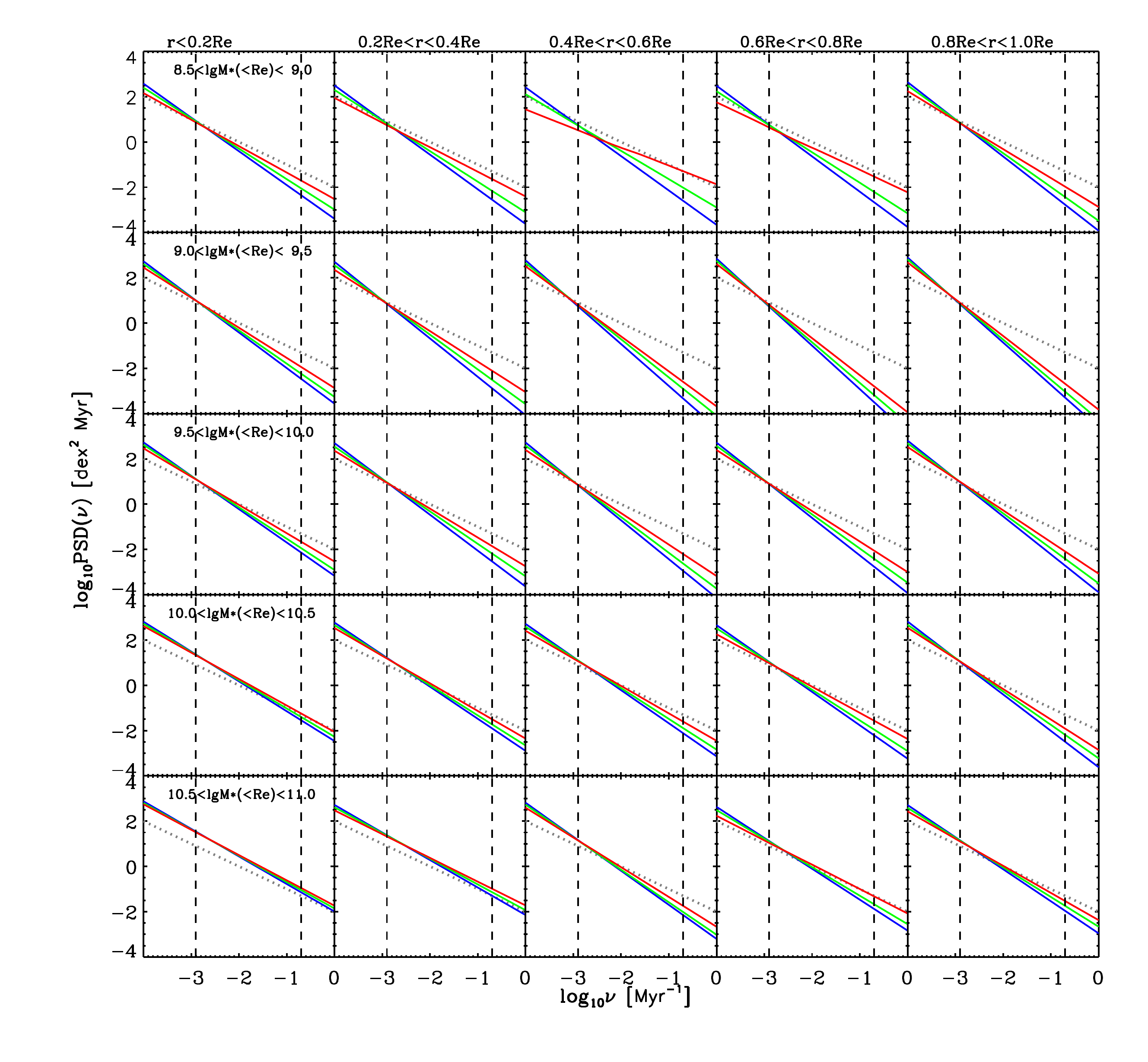}
    \end{center}
  \caption{The returned PSDs of $\Delta$sSFR(t) for the 25 different galactic radius 
  and stellar mass bins shown in Figure \ref{fig:data_resolved}.  The PSDs are displayed with increasing 
  galactic radius from left to right, and with increasing stellar mass from top
  to bottom. At each given galactic radius and stellar mass, the blue, green 
  and red lines show the PSDs obtained by assuming $\sigma_{\rm int}^2=0.0$, $0.25$ 
  and $0.5\sigma_{9}^2$. The two dashed vertical lines indicate  
  the frequencies of 1/5 Myr$^{-1}$ and 1/800 Myr$^{-1}$. The gray dotted lines
  are the same in all panels, for ease of comparison of the panels, and have the same amplitude and $\alpha=1$.}
  \label{fig:psd_3p_resolved}
\end{figure*}

\begin{figure*}
  \begin{center}
    \includegraphics[width=0.8\textwidth]{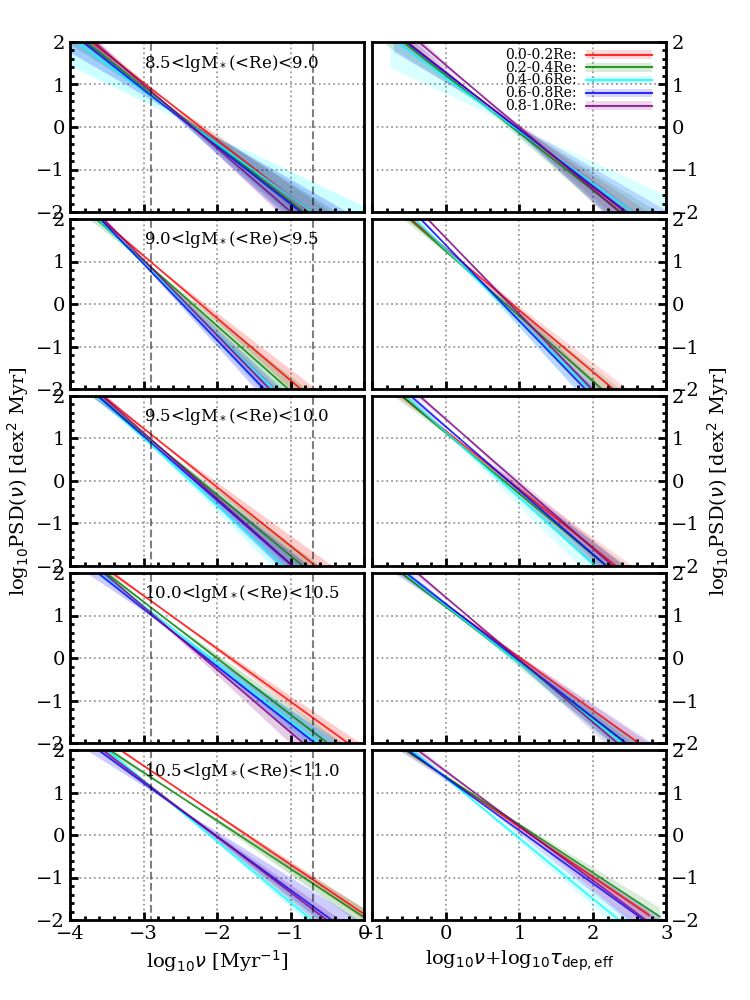}
    \end{center}
  \caption{Left column of panels: the returned PSD($\nu$) from Figure \ref{fig:psd_3p_resolved} over-plotting the five galactic radial bins for each of the five stellar mass bins that are shown top to bottom.  
  In each panel, the PSDs of the five radial bins are displayed 
  in different colors (as denoted in the top right panel of this figure).
  The solid lines show the PSD($\nu$) with $\sigma_{\rm int}^2=0.25\sigma_9^2$, while the shaded regions show the 
  PSD($\nu$) with $\sigma_{\rm int}^2$ from 0.0 to 0.5$\sigma_9^2$.
  The two vertical dashed lines indicates the timescales of 800 Myr and 5 Myr, respectively.
  Right panels: As for the left hand panels, but now shifting the x-axis by the effective gas depletion time, obtained from the stellar surface mass density using the extended Schmidt Law (see text). This shift produces a striking overlap of the PSD. 
   }
  \label{fig:psd_resolved}
\end{figure*}

Following the discussion in Section \ref{subsec:4.3}, we 
only consider cases with $\sigma_{\rm int}^2/\sigma_9<0.5$. 
In this subsection, we will constrain the PSD of $\Delta$sSFR(t) at three distinct $\sigma_{\rm int}$, chosen to be 0.0, 0.5$\sigma_9$ and 
0.7$\sigma_9$. We note that the $\sigma_9$ used here is the $\sigma_9$ of the integrated sample,
rather than the $\sigma_9$ calculated from the bin in question.
%Actually, \citetalias{Wang-19b} proposed that the star formation in galaxies is primarily driven by the global gas inflow, which suggests that the $\sigma_{\rm int}$ is likely the same for the regions at different galactic radii.  

Figure \ref{fig:psd_3p_resolved} shows the PSDs of $\Delta$sSFR(t) that are returned for each of the 
25 panels in Figure \ref{fig:data_resolved}.
In each panel, the blue, green and red lines are the returned PSDs with $\sigma_{\rm int}^2=$0.0, 0.25 and 
0.5$\sigma_9^2$, respectively. 

In each panel, the PSDs returned for the three different $\sigma_{\rm int}$ intersect 
at a point, as was found also for the integrated sample in the previous section. The corresponding timescale (or frequency) of this intersection point varies for the 25 samples considered, but is generally between 
200 Myr and 800 Myr.  Even with the uncertainty introduced by $\sigma_{\rm int}$ there is, within the assumptions of a single power-law PSD, at least one point that is relatively well determined.
%This is in good agreement with the result in Section \ref{subsec:4.3} that 
%adding different amounts of intrinsic scatter causes a rotation of the PSD at a fixed frequency. 

To aid comparison of the panels, the gray dotted lines in Figure \ref{fig:psd_3p_resolved} are the same in all panels and have
$\alpha=1$.  As can be seen, all the returned PSDs are steeper than this, implying that long-timescale variations contribute more (per interval of $\log_{10}\nu$) to the 
variations of $\Delta$sSFR(t) than short-timescale variations. 

To allow direct comparison of the PSD in each of the five galaxy mass bins, we replot them in the left column of panels of Figure \ref{fig:psd_resolved}, 
superposing the regions at different galactic radii onto a single panel.  The lines are color coded by radial distance, and the width represents the range of $\sigma_{\rm int}$ shown in the individual panels of Figure \ref{fig:psd_3p_resolved}.

Careful inspection of Figure \ref{fig:psd_3p_resolved} and Figure \ref{fig:psd_resolved} reveals three interesting results that are developed in the next three sub-sections of the paper. These results will then be discussed in Section \ref{sec:6}. 

\subsection{The connection of the PSD slope with the effective gas depletion time} \label{subsec:5.2}

Comparing the PSDs at different mass bins and at 
different galactic radii, we find that generally speaking the returned PSD appears to flatten with 
both increasing stellar mass at fixed galactic radii, and with decreasing galactic radius
at fixed stellar mass.  The flatter PSD implies larger relative contribution of short timescales compared with long timescales. 

In \citetalias{Wang-19a} and \citetalias{Wang-19b}, we found that the dispersion of $\Sigma_{\rm SFR}$ and SFR79 
decreases with increasing inferred gas depletion time.  We stress once again that the gas depletion 
times used in those papers, and here, are obtained from the extended Schmidt law \citep[e.g.][]{Shi-11, Shi-18}, which 
has the gas depletion time simply proportional to the $\Sigma_*^{-1/2}$.  Consequently, all our statements concerning gas depletion times can be equally well made in terms of $\Sigma_*^{-1/2}$, which reflects a general dynamical time.  It then follows that inferred gas depletion time must therefore decrease with increasing integrated stellar mass and towards the inner regions of galaxies. The effective gas depletion time, $\tau_{\rm dep,eff}$, is defined to be the $\tau_{\rm dep}$  divided by $(1+\lambda)$, where $\lambda$ is the mass-loading factor of the wind outflow for different regions of galaxies,
obtained by fitting the SFR-$\Sigma_*$-metallicity 
relation of MaNGA spaxels under the gas regulator frame (see details given in \citetalias{Wang-19a}).  

\begin{figure}
  \begin{center}
    \includegraphics[width=0.47\textwidth]{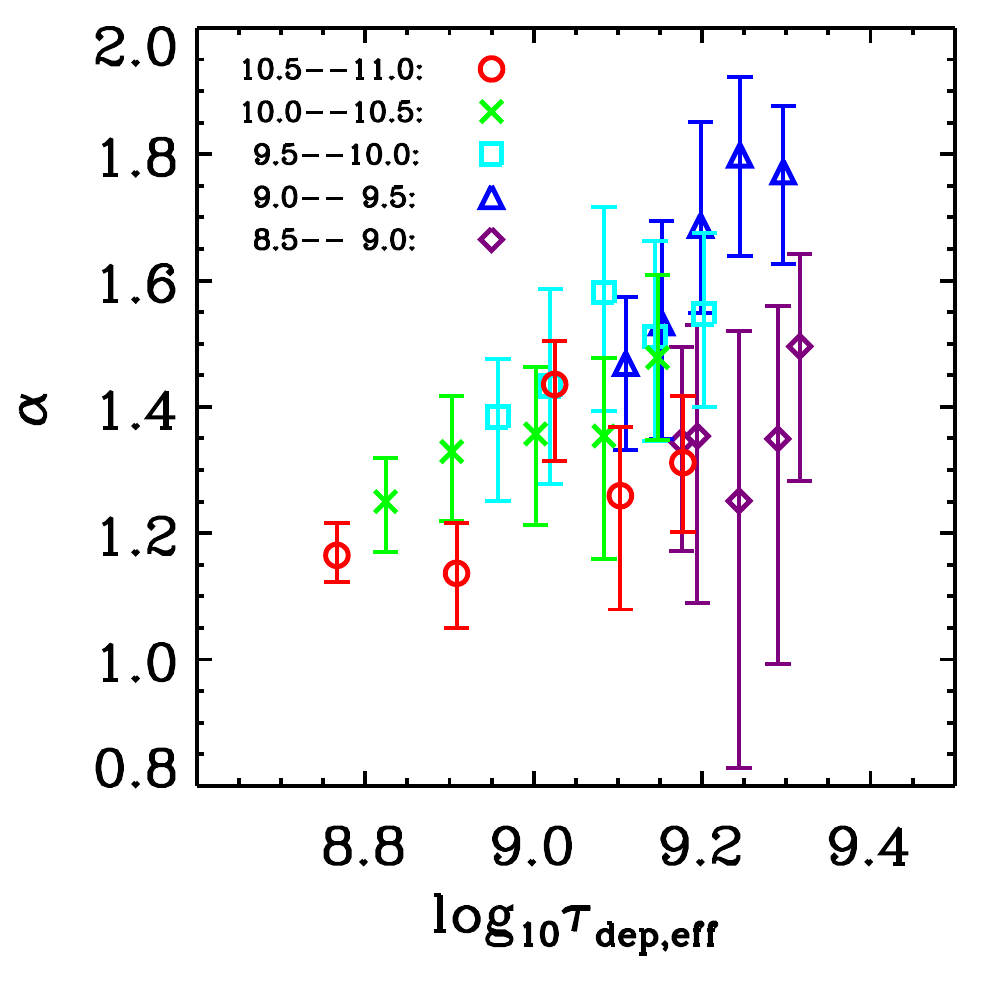}
    \end{center}
  \caption{ The returned $\alpha$ as a function of the inferred effective gas depletion time. The data points are the returned $\alpha$ for $\sigma_{\rm int}^2/\sigma_9^2$=0.25, while the upper limits and the lower limits of the data points are the returned $\alpha$ for $\sigma_{\rm int}^2/\sigma_9^2=$0.0 and 0.5, respectively. 
   The legend for the points is denoted in the top-left corner. For the five points at each stellar mass bin, the $\tau_{\rm dep,eff}$ increases monotonically with radius.
 }
  \label{fig:psd_tdep}
\end{figure}

The trend of $\alpha$ with radius and integrated stellar mass that is visible in Figure \ref{fig:psd_3p_resolved} and Figure \ref{fig:psd_resolved}
suggests that there will be a correlation of $\alpha$ with the $\tau_{\rm dep,eff}$.  This is shown directly in Figure \ref{fig:psd_tdep}.
In Figure \ref{fig:psd_tdep}, the plotted values 
of $\alpha$ are the best-fit values returned by the MCMC for 
$\sigma_{\rm int}^2/\sigma_9^2$=0.25, and the error bars
correspond to the returned $\alpha$ for $\sigma_{\rm int}^2/\sigma_9^2$=0.0 and 0.5.  The points are color-coded for the integrated
stellar mass (since for the five radial points in each mass bin the galactic radius always 
increases monotonically with increasing $\tau_{\rm dep,eff}$). 
As can be seen, the $\alpha$ is indeed a decreasing function of $\tau_{\rm dep,eff}$ 
as a whole.  This again indicates that regions of shorter 
$\tau_{\rm dep,eff}$ have flatter PSD and therefore have relatively more variation 
on shorter timescales. 

\begin{figure}
  \begin{center}
    \includegraphics[width=0.47\textwidth]{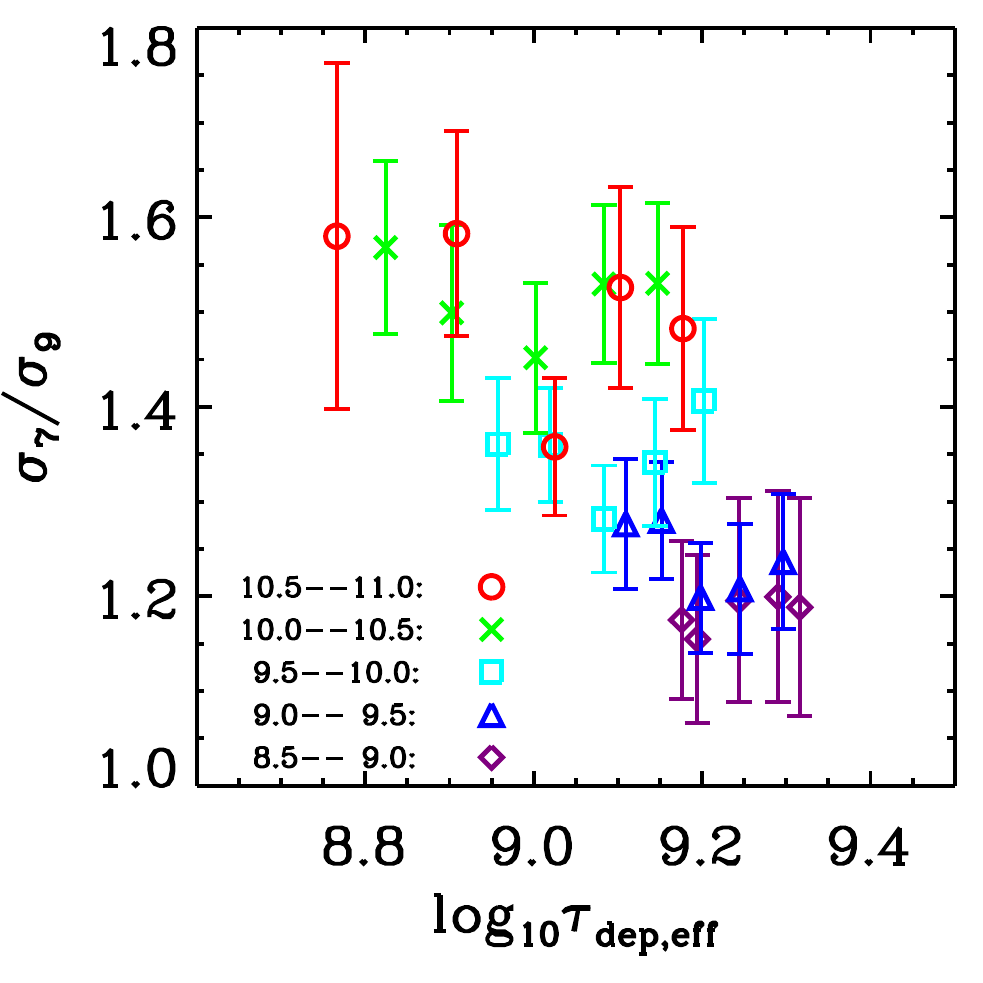}
    \end{center}
  \caption{ As in Figure \ref{fig:psd_tdep}, the observed $\sigma_7/\sigma_9$ is plotted as a function of the inferred effective gas depletion time. The uncertainties of 
  $\sigma_7/\sigma_9$ are computed by the bootstrap method.  
  The legend for the symbols is the same as in Figure \ref{fig:psd_tdep}. 
 }
  \label{fig:sig_tdep}
\end{figure}

As shown in Section \ref{subsec:3.4}, the slope $\alpha$ of
the PSD is primarily determined by the ratio of $\sigma_7$ to $\sigma_9$ (see Figure \ref{fig:para_int}), after the effects of $\sigma_{\rm int}$ on $\sigma_7$ and $\sigma_9$ are removed, if necessary.
The ratio $\sigma_7/\sigma_9$ (possibly corrected for $\sigma_{\rm int}$) is therefore the key to the 
relative distribution of the power on different timescales. 
We therefore plot on Figure \ref{fig:sig_tdep} the directly observed $\sigma_7/\sigma_9$ (not adjusted for any intrinsic scatter) as a function of the inferred 
effective gas depletion times ($\tau_{\rm dep,eff}$) for the five radial regions at five stellar mass bins.
It is clear that the $\sigma_7/\sigma_9$ ratio decreases with increasing effective 
gas depletion time. We note that this empirical result is derived purely from the observations and is therefore quite independent of the assumptions made about
the form of the PSDs in the present work. 

%However, interpreting this result under our frame, regions with 
%shorter $\tau_{\rm dep,eff}$ would have flatter PSD of $\Delta$sSFR(t),  
%assuming the intrinsic scatter is not dominant. This further means that 
%regions of shorter $\tau_{\rm dep,eff}$ tend to have relative more variations 
%at short timescale.  

We note that in Figure \ref{fig:psd_tdep} (and Figure \ref{fig:sig_tdep}), there is not a 
clear correlation between $\sigma_7/\sigma_9$ (or $\alpha$) with the $\tau_{\rm dep,eff}$ at the lowest stellar mass bin represented by purple points.   This may simply be due to the narrow range of $\tau_{\rm dep,eff}$ within that mass bin.

\subsection{The connection of the PSD amplitude with effective gas depletion time } \label{subsec:5.3}

\begin{figure*}
  \begin{center}
    \includegraphics[width=0.8\textwidth]{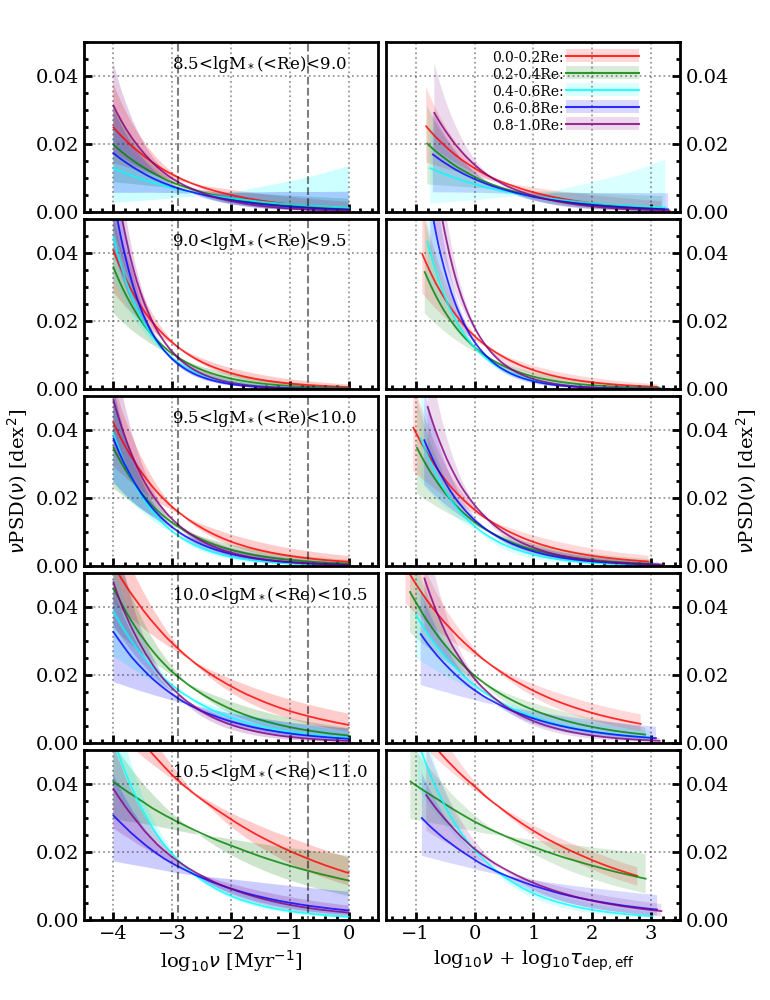}
    \end{center}
  \caption{ The same as Figure \ref{fig:psd_resolved} but now showing the $\nu$PSD($\nu$) function, which is here plotted with a linear scale. 
  It is the $\nu$PSD($\nu$) function, rather than PSD($\nu$), which shows the contributions to the variance per interval of $\log \nu$ at different frequencies.   
 }
  \label{fig:psd_shift}
\end{figure*}

The amplitude of the PSD is linked to the overall variation of star-formation rates on a given timescale. Specifically, the $\sigma_7^2$ and $\sigma_9^2$ will be given by an integral of $\nu {\rm PSD}(\nu)d \log \nu$ from the inverse Hubble time up to the frequencies corresponding to 5 Myr or 800 Myr timescales (see Equation \ref{eq:5}).   Accordingly we replot the PSD that were presented in the left hand panels of Figure \ref{fig:psd_resolved} as the linear $\nu$PSD($\nu$) vs. log $\nu$ in the equivalent left-hand panels of Figure \ref{fig:psd_shift}. Integration of these curves to the left of the two dashed vertical lines will give a measure of the variation of star-formation rates sampled on these two timescales, i.e. what we have called $\sigma_7^2$ and $\sigma_9^2$. As one can see, for all the returned PSDs, 
the contribution of the power per $\log_{10}\nu$ decreases with the frequency, but the integrals up to a given frequency, are larger with increasing stellar mass and towards inner regions of galaxies, i.e. varying with the inferred $\tau_{\rm dep,eff}$. 

This change in amplitude is to be expected from the results in \citetalias{Wang-19b}, where we showed (see the figure 15) that the observed dispersion in $\Sigma_{\rm SFR}$ increases with decreasing $\tau_{\rm dep,eff}$.  While this agreement is ultimately to be expected, since the same MaNGA data was used in the two analyses, although it should be noted that the MCMC analysis returning the PSD used the full two-dimensional distributions of points in the 25 panels of Figure \ref{fig:data_resolved}, whereas the analysis in \citetalias{Wang-19b} was based only on the computed dispersions of each of these quantities.

\subsection{The convergence of the different PSDs when shifted by the effective gas depletion time} \label{subsec:5.4}

\begin{figure*}
  \begin{center}
    \includegraphics[width=0.45\textwidth]{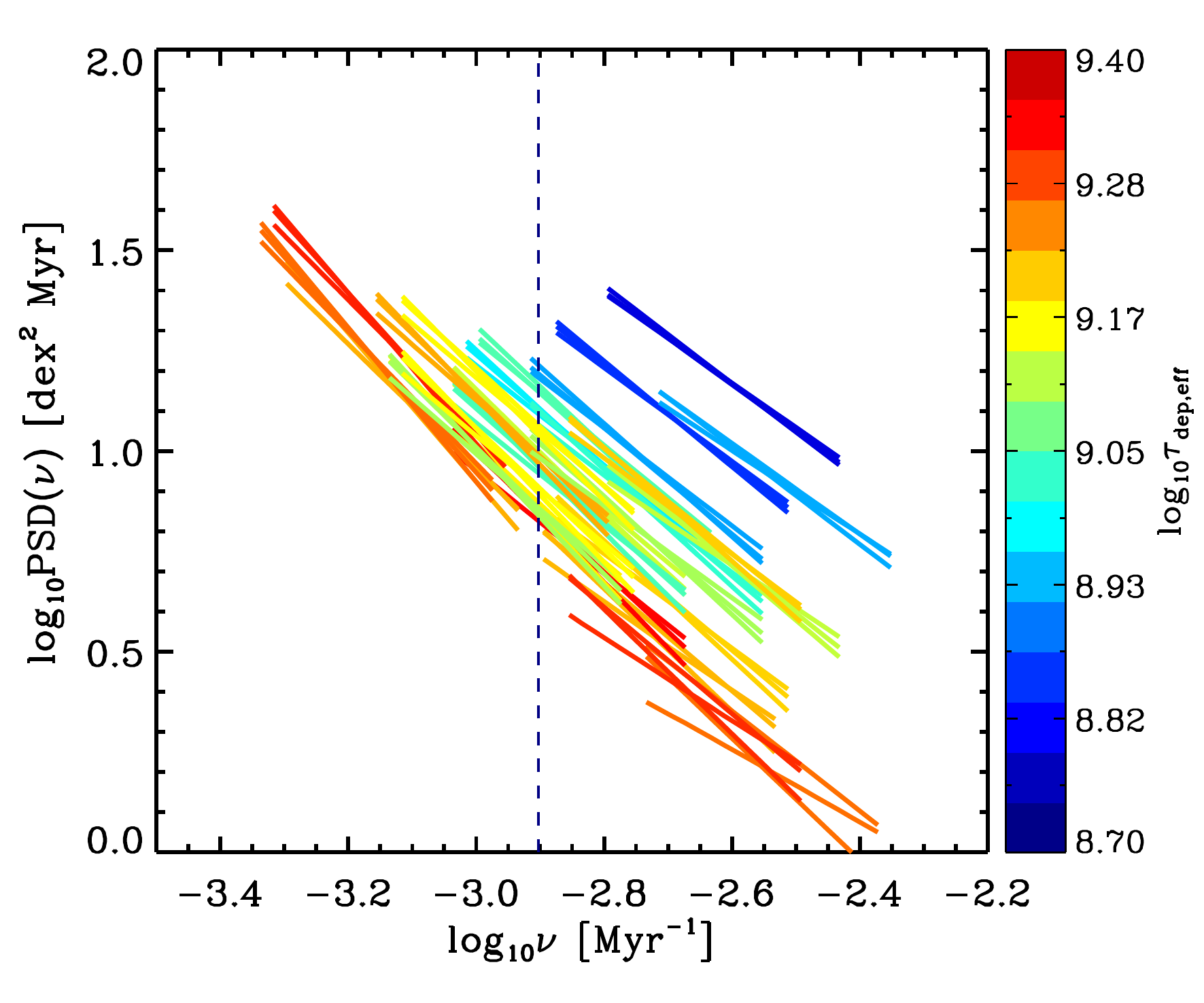}
    \includegraphics[width=0.45\textwidth]{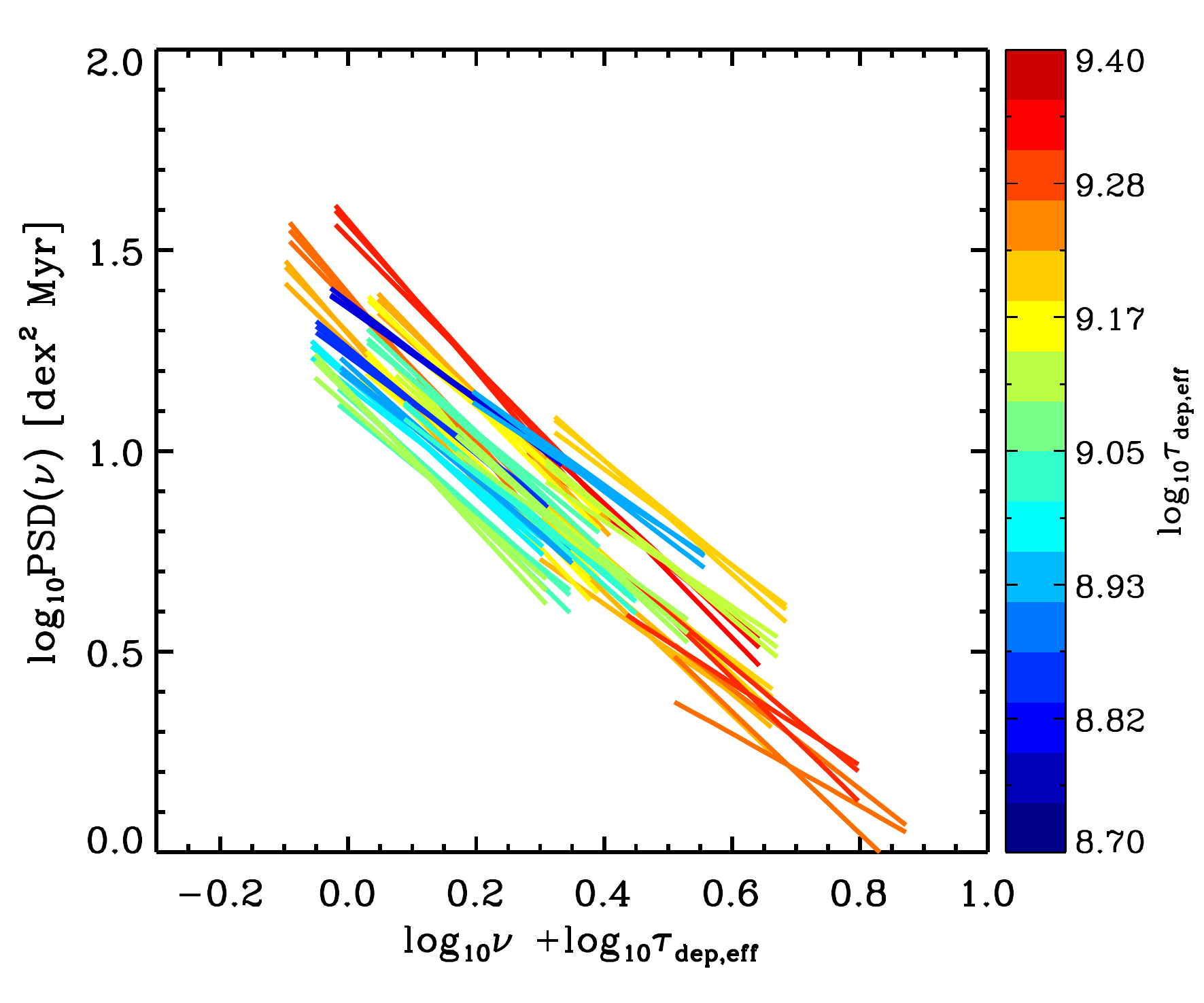}
    \end{center}
  \caption{ Left panel: The returned PSDs from the five left hand panels of Figure \ref{fig:psd_resolved} extracted near the pivot points (within $\pm$0.2 dex), where the PSDs are nearly independent of the assumed $\sigma_{\rm int}$.  The different colors indicate the different effective gas depletion times according to the color scale on the right. 
  The dashed vertical line indicates the frequency of 1/800 Myr$^{-1}$. 
  Right panel: The same as the left panel but now the PSD are shifted in frequency by the inferred effective gas depletion time, as in the right hand column of panels in Figure \ref{fig:psd_resolved}.  This panel shows the remarkably constant PSD that is obtained, independent of radius and integrated mass, once the PSD are shifted by the effective depletion time.
 }
  \label{fig:psd_pivot}
\end{figure*}

The foregoing two sub-sections have emphasized the connection between the returned PSD and the inferred effective gas depletion timescale (or, as repeatedly stressed, any other quantity determined by the stellar surface mass density).
To explore this connection further, we now normalize the frequency axes of the left panels of both Figure \ref{fig:psd_resolved} and Figure \ref{fig:psd_shift} by shifting the PSD curves in frequency by the inferred effective gas depletion time, and replot them in the right hand column of panels in each Figure.  

An intriguing result emerges in that the PSDs of
Figure \ref{fig:psd_resolved} are now noticeably more tightly overlapped for the regions of different galactic radii, 
and also for galaxies of different mass, especially for the three highest mass bins.  This suggests that there might be a 
universal PSD($\nu$), valid at all radii in all the galaxies, at a fixed frequency {\it relative} to a $\nu_{\rm dep,eff}$, defined as the inverse of $\tau_{\rm dep,eff}$.  We show this result in a different way on Figure \ref{fig:psd_pivot}, where we plot on a single diagram short segments of all of the 25 PSDs, extracted from around the ``pivot point'' of convergence when $\sigma_{\rm int}$ is varied, since this is where we may consider the PSD to be best determined in our analysis.  These PSD segments are then plotted without (left) and with (right) the frequency shift using the inferred effective depletion time.  Clearly the shift in frequencies produces a much tighter overlap of the PSD from the 25 different samples of integrated stellar mass and relative radius.

%%%%%%%
While intriguing, one should not over-interpret this result.  Not least, we noted earlier that the slopes of the PSD 
at different galactic radii are slightly different, as noted above, although this does depend in detail on the assumed $\sigma_{\rm int}$. 
Furthermore, even if the shifted PSD overlap, it does not mean that the shifted $\nu$PSD functions will overlap, in fact the converse.  
As noted, it 
is the $\nu$PSD function which gives the contribution to the variance from each $\log_{10}\nu$ interval.  We will return to this point below In Section 6.2.

%and it is therefore this that might be expected to be ``universal" under some circumstances.  if we had stumbled across some new %law of star-formation in galaxies. As can be seen in the righthand panels of 
%Figure \ref{fig:psd_shift}, significant differences in
%$\nu$PSD are still seen after the frequency normalization. 

%3, The possible explanation ... based on Wang-19, and the predictions ...

We stress that the results presented in Figure \ref{fig:psd_tdep},
\ref{fig:psd_shift} and \ref{fig:psd_pivot} are derived directly from the 
PSD extracted from the observational quantities shown in Figure \ref{fig:data_resolved} . The only assumptions involved are of the power-law nature of the PSD, and the choice of the extended Schmidt Law to infer the gas depletion timescale from the stellar surface density.
%The process of obtaining these
%results has nothing with the dynamical gas regulator model we 
%proposed in \citetalias{Wang-19a}, 
In the next section we will present 
a discussion of the possible explanation of these results.   
%In a word, one should take these results as the observational results, 
%rather than model predictions. 

\section{Discussion} \label{sec:6}

\subsection{Caveats} \label{subsec:6.1}

The major assumption of this work is that of ergodicity, i.e.
that (i) that the variations in the population at a given epoch are produced only of temporal variations in individual galaxies and (ii) that the sample galaxies in each case have the same non-evolving PSD of the $\Delta$sSFR(t). We already explained that the second is for instance clearly not strictly the case for the integrated sample in Figure \ref{fig:global} because of the different forms of the separate panels in Figure \ref{fig:data_resolved}.  It is possible that each panel in Figure \ref{fig:data_resolved} might in turn be found to be break down into different subsets of galaxies, with different distributions of points in the $\Delta$sSFR7-$\Delta$sSFR9 diagram.  

We have considered in some detail in the paper how we could treat violations of the first assumption by 
introducing the concept of intrinsic scatter $\sigma_{\rm int}$ of the SFMS.  Needless to say, 
the case in the real Universe may be more complicated in that galaxies 
can break the assumption of ergodicity in different ways. 

In extracting the PSD, we assume that the PSD of galaxies 
has not evolved significantly over the last billion years.
It would not be surprising if the PSD of $\Delta$sSFR at high redshift was different 
from the local ones at much earlier times, although the similar dispersions of the SFMS suggests that any such variations are probably not very large. This could be examined observationally by applying the same methods in \citetalias{Wang-19b} and the present work 
to high-z galaxies, which we plan to do in the future.   We note however that variations of the PSD$(\nu)$ on timescales much longer than 1 Gyr should be irrelevant in the present analysis.
%(for $\nu > 1$ Gyr$^{-1}$ or mimic the effects of $\sigma_{\rm int}$ (for $\nu > 1$ Gyr$^{-1}$.

%because of their much higher inflow rates and much higher gas densities and generally represent a 
%very different star-forming environment to that in the local Universe. 

%We emphasize that the PSD obtained in the present work is the PSD of $\Delta$sSFR(t)
%at low-redshift Universe. 

In addition, the shape of the PSD of $\Delta$sSFR(t) in the real Universe may well not be as simple as the single power-law function that we assumed. 
However, we only have the information of the averaged SFR on
two timescales, which clearly restricts exploration of more complicated forms of the PSD. In addition, as we have seen, the slope of the power-law PSD is highly degenerate with the (unknown) ``intrinsic scatter". 
More SFR indicators with well understood accuracy are needed to break this degeneracy
in the future.  
Since the intrinsic scatter cannot be constrained in the present work, the main results 
have been presented assuming $\sigma_{\rm int}^2/\sigma_9^2<0.5$. Although this assumption 
is consistent with the results from hydro-dynamical simulations, direct observational 
confirmation is still lacking. 

Finally, in modelling the PSD, we have considered the uncertainties of galaxies on the 
$\Delta$sSFR7-$\Delta$sSFR9 diagram, assuming that the uncertainties of 
$\Delta$sSFR7 and $\Delta$sSFR9 are taken only from SFR7 and SFR9, i.e. ignoring any uncertainty in the observational measurement of 
stellar mass (or stellar surface mass density).  As noted above, errors in the measurement of the (current) 
stellar mass would perturb $\Delta$sSFR7 and $\Delta$sSFR9 equally, and therefore have the same effect as our ''intrinsic scatter". Therefore,
this source of uncertainty should have been automatically been taken into account, but this alone shows that any assumption of $\sigma_{\rm int} = 0$ is unlikely to be valid.   

\subsection{Interpretation: a possible explanation within the gas regulator model} \label{subsec:6.2}

In this work, we have found connections between the PSDs of $\Delta$sSFR(t) with 
the effective gas depletion time: (i) regions with shorter $\tau_{\rm dep,eff}$
show larger integrated power, and {\it vice versa}, (ii) regions with shorter $\tau_{\rm dep,eff}$
have a more significant contribution of the power at short timescales because of their flatter $\alpha$, and (iii) when the PSDs are shifted by an amount corresponding to the $\tau_{\rm dep,eff}$, the PSDs appear to overlap to a rather striking degree, independent of the relative radius of the region or the integrated stellar mass of the galaxy.

The first statement is in good agreement with the result in \citetalias{Wang-19a} and \citetalias{Wang-19b}
that the dispersion of $\Sigma_{\rm SFR}$ (or SFR79) decreases with increasing $\tau_{\rm dep,eff}$ and is in effect a restatement of that result.  While these three analyses all used the same input data, this nevertheless 
provides a good consistency check for the methodology of determining the PSDs.  As 
proposed in \citetalias{Wang-19a}, this result can be understood in the framework of the dynamical gas regulator model. 

In \citetalias{Wang-19a}, we saw what would happen if we drove the gas regulator system of \cite{Lilly-13} with a periodically varying inflow.  As would be expected, the system then has an SFR response that   
is also
periodic with an amplitude (and small phase shift) that depends on the ratio of the driving period to the effective gas depletion timescale.   It was found that this 
can reproduce the observed correlation between the dispersion of $\Sigma_{\rm SFR}$ and 
$\tau_{\rm dep,eff}$ well. 

The basic assumption in this interpretation is that the variation 
of SFR (or $\Sigma_{\rm SFR}$) from galaxy to galaxy is the response to externally driven temporal variations of the gas inflow rate, rather than intrinsic (time-independent) differences between galaxies. This was confirmed in \citetalias{Wang-19b}, where
we found that the regions with shorter $\tau_{\rm dep,eff}$ indeed do show larger {\it temporal} variations of the SFR, as evidenced by the SFR change parameter SFR79.

In the analysis of \citetalias{Wang-19a}, driving the gas regulator system with a purely sinusoidal 
inflow rate, produces a sinusoidal response SFR. The amplitude of the variations in SFR is reduced from the amplitude of variations in the inflow rate by 
a frequency dependent response curve.  In this ideal case, the response curve can be written analytically as
\begin{equation} \label{eq:8}
     f = \frac{1}{[1+(2\pi\tau_{\rm dep,eff}/\tau_{\rm P})^2]^{1/2}},  
\end{equation}
where $\tau_{\rm P}$ is the period of the gas inflow rate. 
According to Equation \ref{eq:8},
the response curve increases with $\tau_{\rm P}$ at given $\tau_{\rm dep,eff}$: $f\sim1$ at 
$\tau_{\rm P}\gg\tau_{\rm dep,eff}$ and 
$f\sim0$ at $\tau_{\rm P}\ll\tau_{\rm dep,eff}$. 
This means that long-timescale variations ($\gg\tau_{\rm dep,eff}$) in the inflow 
rate are fully passed on to the resulting SFR through the gas regulator system, while
short-timescale ($\ll\tau_{\rm dep,eff}$) variations in the inflow rate are eradicated. 

%For a given variation of the inflow rate,
%shorter $\tau_{\rm dep,eff}$ would lead to larger overall variations 
%of the SFR,
%
%XXX SJL  GETTING CONFUSED
%
%and more relative contribution of the variations at short timescale,

%In the context of the PSD, Equation 8 would predict that the PSD of the $\Delta sSFR(t)$ would be equal to the PSD of the (specific) inflow multiplied by the response curve ${[1+(2\pi\tau_{\rm dep,eff}/\tau_{\rm P})^2]^{1/2}}^{-1}$.
%according to Equation \ref{eq:8}.  
%Therefore, assuming that regions at different galactic radii experience %similar variations in 
%the inflow rate, the results in Figure \ref{fig:sig_tdep} and Figure %\ref{fig:psd_tdep}
%are qualitatively expected in the framework of the gas regulator model. 

\citetalias{Wang-19a} suggested that the PSD of the SFR response of the regulator should therefore be the PSD of the inflow rate
multiplied by the $f^2$ response curve given in Equation \ref{eq:8}. 
%This then predicts the connection between the PSDs of the SFR for different %effective 
%gas depletion times. 

We do not know the PSD of the inflow rate.  But, if we assume a flat PSD for this ($\alpha=0$), and assume that this is the same for all of the 25 regions considered (we stress that both are rather large assumptions) then we would find that 
the PSDs of the resulting SFR would be exactly overlapped when one scaled the frequency with the $\tau_{\rm dep,eff}$, as
was done in Figures \ref{fig:psd_resolved} and \ref{fig:psd_pivot}. We suggest that this may conceivably be the explanation for  
the seemingly uniform PSD when we scale with the frequency with $\tau_{\rm dep,eff}$.  Furthermore, under these assumptions one would predict a slope of the PSD of $\Delta$sSFR(t) to have $\alpha \sim 2$.  

More generally, the fact that the different PSD of $\Delta$sSFR(t) overlap when shifted by a physically meaningful amount may be the signature of systems (differentiated only by their characteristic response frequency ${\tau_{\rm dep,eff}}^{-1}$) responding to uniform {\it external} drivers, rather than being driven by an entirely {\it internal} process. In the latter case one might expect that the power per log frequency interval, i.e. $\nu$PSD($\nu$), would be the constant quantity as some characteristic timescale changed.  However, these considerations are completely speculative at this point, especially given our very limited theoretical understanding of the PSD of inflow and our still very limited observational constraints on the PSD of $\Delta$sSFR(t).

\section{Summary and Conclusion} \label{sec:7}
\label{sec:conclusion}

In this work, we have attempted to constrain the PSD of $\Delta$sSFR(t) for Main Sequence galaxies (the logarithmic offset of each galaxy from the median Main Sequence), and for radial regions within them. 
This has been based on the star formation change parameter, SFR79, introduced and calibrated in \citetalias{Wang-19b}.  
As we discussed in \citetalias{Wang-19b}, the scatter of SFR79 contains the information about 
the time variability of the SFR.

We take the advantage of a well-defined SF galaxy sample
from the MaNGA survey \citep{Wang-18a}, which contains nearly 1000 galaxies. 
While we do a full MCMC fit to the observed distribution of points in the $\Delta$sSFR7-$\Delta$sSFR9 plane, the main observational information that is used to
constrain the PSD of $\Delta$sSFR(t) is effectively the dispersion of the SFMS on different timescales ($\sigma_7$ and $\sigma_9$) 
and the dispersion of the SFR79 ($\sigma_{79}$) for individual galaxies. 

In constructing the PSD, we would ideally be able to assume that $\Delta$sSFR(t), is a stochastic variable that exhibits both
stationarity and ergodicity, i.e. that the distribution of $\Delta$sSFR is constant and that the variation of sSFR across the population at a given epoch reflects only the temporal variations in individual objects (see the extensive discussion in Section \ref{sec:2}). 

While stationarity is a reasonable assumption, it may well be that galaxies violate ergodicity.  Subsets of individual galaxies could well lie 
systematically above (or below) the SFMS throughout their lifetimes.  A significant non-ergodic contribution to the dispersion of the SFMS could also come from observational errors in the determination of the (present-day) masses of galaxies. 

We have considered this likely
non-ergodicity by introducing a parameter, the intrinsic scatter of the SFMS, $\sigma_{\rm int}$, which represents any contribution to the dispersion in sSFR that is not due to temporal variations of individual objects.
The intrinsic scatter must be less than $\sigma_9$, the observed dispersion of the SFMS when the SFR is averaged over the last Gyr. 

The effects of $\sigma_{\rm int}$ could in principle be accommodated by introducing Fourier modes at very low frequencies, i.e. $\nu \ll \tau_{H}^{-1}$. However, we disfavour this approach because there is no reason to suppose that the causes of $\sigma_{\rm int}$ should in any way be related to the amplitudes of modes in the physically meaningful part of the PSD at $\nu > \tau_{H}^{-1}$.

Assuming a single power-law for the PSD of $\Delta$sSFR(t), truncated at $\nu \sim$ 0.1 Gyr$^{-1}$, we use an MCMC approach to explore the parameter space of permitted PSD by matching the distribution of galaxies
on the $\Delta$sSFR7-$\Delta$sSFR9 diagram. 

We examine the methodology by applying it first to an ``integrated" sample in which the $\Delta$sSFR7 and $\Delta$sSFR9 are computed for an entire galaxy within \re, and in which galaxies of all stellar masses are considered together (even though we know that this sample is not strictly ergodic).
We find that the returned slope of PSD is tightly degenerate with the assumed value of the intrinsic scatter, the slope of PSD becoming flatter as the $\sigma_{\rm int}$ increases.
Ignoring the intrinsic scatter would undoubtedly overestimate 
the slope of PSD \citep[c.f.][]{Caplar-19, Hahn-19}. 

%The $\sigma_{79}$ does not depend on the $\sigma_{\rm int}$, therefore the 
%constrained $P_{\nu_0}$ and $\alpha$ move on the parameter space following 
%the constant $\sigma_{79}$ at different $\sigma_{\rm int}$ (see Figure %\ref{fig:psd_2p_3p}). 
%This suggests the importance of 
%SFR79 calibrated in \citetalias{Wang-19b} in determination of the PSD of %$\Delta$sSFR(t). 

The slope of the PSD for this integrated sample varies from 1.2 to 1.5 at 
$\sigma_{\rm int}^2<0.5\sigma_9^2$, indicating that the power contributed (per
interval of $\log_{10}\nu$) by 
long-timescale variations is always larger than that contributed by the short-timescale variations.  In addition, the PSDs intersect at a point at a timescale of $\sim$ 400 Myr (see Figure \ref{fig:psd_2p_3p}), independent of the value of $\sigma_{\rm int}$ within the above constraints.

In principle, we cannot constrain the $\sigma_{\rm int}$ based on the data 
we have in the present work. However, the hydro-dynamical simulations suggest that
$\sigma_{\rm int}$ is not close to $\sigma_9$, because significant variations of
$\Delta$sSFR on timescales between 1 Gyr to 10 Gyr are seen \citep{Matthee-19}.
Therefore, in the present work, we present further results assuming that
$\sigma_{\rm int}^2<0.5\sigma_9^2$.

We then apply this methodology to a spatially-resolved dataset to obtain the main observational result of the paper: constraints on the 
PSDs of $\Delta$sSFR(t) for 25 regions obtained by considering five annular radial bins for galaxies in five different stellar mass bins.
The main results of this spatially resolved analysis are as follows: 

\begin{itemize}

\item The slope of the PSD for the spatially-resolved data varies from 1.0 to 2.0 
for $\sigma_{\rm int}^2<0.5\sigma_9^2$. As for the integrated sample,
the PSDs for a given radius-mass bin intersect at a point as $\sigma_{\rm int}$ is varied (within the above range).
The corresponding timescale of this pivot point varies from 200 Myr to 800 Myr 
for regions of different galactic radii at different stellar mass bins (see Figure \ref{fig:psd_3p_resolved}). 

\item  There is a strong correlation between the returned PSDs and the inferred effective
gas depletion times, where the latter are estimated using the extended Schmidt Law from the observed stellar surface mass density plus an estimate of mass loss from winds.  The regions with shorter $\tau_{\rm dep,eff}$
show both larger integrated power, and a shallower $\alpha$, i.e. they have more significant power (but still not dominant) on short timescales (see Figure \ref{fig:psd_resolved}, Figure \ref{fig:psd_tdep} and Figure \ref{fig:psd_shift}).  

\item  Intriguingly, if we scale the frequency with $\tau_{\rm dep,eff}$, the 25 different returned PSDs largely overlap, especially around the pivot points where the PSD is arguably best determined. 
 This suggests a remarkably uniform PSD($\nu$) as a function of $\log_{10}\nu/\nu_{\rm dep,eff}$, where $\nu_{\rm dep,eff}$ is the inverse of
 $\tau_{\rm dep,eff}$ (see Figure \ref{fig:psd_resolved} and Figure \ref{fig:psd_pivot}). 

\end{itemize}

We emphasize that the above results are obtained from the observations.  The apparent connections with the inferred $\tau_{\rm dep,eff}$ however continue to be in
conceptual agreement 
with what is expected when a gas regulator system \citep{Lilly-13} responds to variations in the inflow rate, which we 
explored in \citetalias{Wang-19a}.   Not least, if such gas regulator systems (which will differ only in their response frequency, set by the depletion time) were all to be fed by the same time-varying inflow, with a flat $\alpha \sim 0$ PSD, then the PSD of the SFR response would indeed appear to be the same when shifted in frequency by the depletion time, as observed (see Figure \ref{fig:psd_pivot}).  Whether this is indeed the explanation for the intriguing results presented in this paper remains to be determined.
%Assuming a uniform time-varying inflow rate,
%shorter $\tau_{\rm dep,eff}$ would lead to larger overall variation 
%of resulting SFR, and more relative contribution of the variations at 
%short timescale through the gas regulator system. 

\acknowledgments

Funding for the Sloan Digital Sky Survey IV has been provided by
the Alfred P. Sloan Foundation, the U.S. Department of Energy Office of
Science, and the Participating Institutions. SDSS-IV acknowledges
support and resources from the Center for High-Performance Computing at
the University of Utah. The SDSS web site is www.sdss.org.

SDSS-IV is managed by the Astrophysical Research Consortium for the
Participating Institutions of the SDSS Collaboration including the
Brazilian Participation Group, the Carnegie Institution for Science,
Carnegie Mellon University, the Chilean Participation Group, the French Participation Group, 
Harvard-Smithsonian Center for Astrophysics,
Instituto de Astrof\'isica de Canarias, The Johns Hopkins University,
Kavli Institute for the Physics and Mathematics of the Universe (IPMU) /
University of Tokyo, Lawrence Berkeley National Laboratory,
Leibniz Institut f\"ur Astrophysik Potsdam (AIP),
Max-Planck-Institut f\"ur Astronomie (MPIA Heidelberg),
Max-Planck-Institut f\"ur Astrophysik (MPA Garching),
Max-Planck-Institut f\"ur Extraterrestrische Physik (MPE),
National Astronomical Observatory of China, New Mexico State University,
New York University, University of Notre Dame,
Observat\'ario Nacional / MCTI, The Ohio State University,
Pennsylvania State University, Shanghai Astronomical Observatory,
United Kingdom Participation Group,
Universidad Nacional Aut\'onoma de M\'exico, University of Arizona,
University of Colorado Boulder, University of Oxford, University of Portsmouth,
University of Utah, University of Virginia, University of Washington, University of Wisconsin,
Vanderbilt University, and Yale University.

%This research has been supported by the Swiss National Science Foundation. 

%\bibliography{rewritebib2.bib}
\bibliography{ms.bbl}

\clearpage
%\appendix

%%%%%%%%%%%%The End%%%%%%%%%%%%%%%%%%%%%%%%%%%%%%%%%%%%%%%%%
\label{lastpage}
\end{document}